\documentclass[12pt,preprint]{aastex}

\slugcomment{Submitted to the Astrophysical Journal.}

\shorttitle{The Physical Properties of $\delta$ Orionis}
\shortauthors{Harvin et al.}

\begin{document}

\title{Tomographic Separation of Composite Spectra. VIII. \\
The Physical Properties of the Massive Compact Binary in the 
Triple Star System HD 36486 ($\delta$ Orionis A)}

\author{James A. Harvin, Douglas R. Gies\altaffilmark{1}, William G. Bagnuolo, Jr.}
\affil{Center for High Angular Resolution Astronomy, \\
    Department of Physics and Astronomy, \\
    Georgia State University, \\
    Atlanta, GA 30303}
\email{harvin@chara.gsu.edu, gies@chara.gsu.edu, bagnuolo@chara.gsu.edu}

\author{Laura R. Penny}
\affil{Department of Physics and Astronomy, \\
    College of Charleston, \\
    Charleston, SC 29424}
\email{pennyl@cofc.edu}

\and

\author{Michelle L. Thaller\altaffilmark{1,2}}
\affil{Infrared Processing and Analysis Center, \\
    California Institute of Technology and Jet Propulsion Laboratory, \\
    Pasadena, CA 91125}
\email{thaller@ipac.caltech.edu}

\altaffiltext{1}{Visiting Astronomer, Kitt Peak National Observatory, National Optical
Astronomy Observatories, operated by the Association of Universities for Research in Astronomy,
Inc., under contract with the National Science Foundation.}

\altaffiltext{2}{Visiting Astronomer, Mount Stromlo and Siding Springs Observatories,
Australian National University}

\begin{abstract}
We present the first double-lined spectroscopic orbital elements 
for the central binary in the massive triple, \mbox{$\delta$~Orionis~A}.
The solutions are based on 
fits of cross-correlation functions of {\it IUE} high dispersion UV 
spectra and \ion{He}{1} $\lambda 6678$ profiles. 
The orbital elements for the primary agree well with 
previous results, and in particular, we confirm the apsidal 
advance with a period of $224.5 \pm 4.5 {\rm ~y}$.   
We also present tomographic reconstructions of the primary and secondary stars'
spectra that confirm the O9.5~II classification of the primary and indicate a 
B0.5~III type for the secondary.  The relative line strengths between the 
reconstructed spectra suggest magnitude differences of 
$\Delta m = -2.5 \log (F_s/F_p) = 2.6 \pm 0.2$ in the UV and 
$\Delta m = 2.5 \pm 0.3$ at 6678~\AA .   The widths of the 
UV cross-correlation functions are used to estimate the 
projected rotational velocities, $V\sin i = 157 \pm 6 {\rm ~km~s}^{-1}$
and $138 \pm 16 {\rm ~km~s}^{-1}$ for the primary and secondary, respectively 
(which implies that the secondary rotates faster than the orbital motion).  

We used the spectroscopic results to make a constrained fit of 
the {\it Hipparcos} light curve of this eclipsing binary, and 
the model fits limit the inclination to the range $i = 67^\circ - 77^\circ$. 
The lower limit corresponds to a near Roche-filling configuration 
that has an absolute magnitude which is consistent with the photometrically determined 
distance to Ori~OB1b, the Orion Belt cluster in which \mbox{$\delta$~Ori} resides. 
The $i = 67^\circ$ solution results in masses of $M_p=11.2~M_\odot$ and 
$M_s = 5.6~M_\odot$, both of which are substantially below  
the expected masses for stars of their luminosity.   
The binary may have experienced a mass ratio reversal 
caused by Case~A Roche lobe overflow, 
or the system may have suffered extensive mass loss 
through a binary interaction (perhaps during a common envelope phase) 
in which most of the primary's mass was lost from the system rather than 
transferred to the secondary. 

We also made three component reconstructions to search for the 
presumed stationary spectrum of the close visual companion, \mbox{$\delta$~Ori~Ab}
\mbox{(Hei~42~Ab)}.   There is no indication of the spectral lines of this
tertiary in the UV spectrum, but a broad and shallow feature is 
apparent in the reconstruction of \ion{He}{1} $\lambda 6678$
indicative of an early B-type star.  The tertiary may be a rapid rotator 
($V\sin i \approx 300 {\rm ~km~s}^{-1}$) or a spectroscopic binary. 
\end{abstract}



\section{Introduction}

In this series of papers we have applied a 
version of the Doppler tomography algorithm \citep{bag94} 
to reconstruct the component ultraviolet and optical spectra
of massive close binary systems, 
and we have investigated their spectral morphology, 
rotational velocities, and relative flux contributions. 
We were successful in three component spectral reconstructions of 
the triple systems 55~UMa \citep{liu97} and $\delta$~Cir \citep{pen01}, and 
we are continuing to examine the ultraviolet spectra of several
OB systems known to be triples from speckle interferometry \citep{mas98}.
Here we turn our attention to the most famous of this 
group, and one of the brightest O-stars in the sky, $\delta$~Orionis. 

The visual and spectroscopic components of the $\delta$ Orionis system
(Mintaka, 34~Ori, HD~36486, HR~1852, \mbox{BD--00~983}, ADS~4134, Hei~42, HIP~25930)
are depicted in Fig~1.
The Washington Double Star Catalog \citep{wor97} lists two distant 
visual companions: $\delta$~Ori~B \mbox{(BD--00~983B)}, 
a 14.0 mag star 33$\arcsec$ away, and 
$\delta$~Ori~C (HD~36485), a 6.85 mag star 
at a separation of 53$\arcsec$.
Both of these targets 
are far outside the spectroscopic apertures used on
the {\it International Ultraviolet Explorer Satellite (IUE)},
and so will not be considered further here.   
However, the remaining visual companion, $\delta$~Ori~Ab
(Hei~42~Ab), is very close, and its light will contribute 
to the spectra we discuss below.  
This companion was discovered by \citet{hei80} in 1978 
at a separation of $0\farcs2$, and subsequent speckle interferometry
\citep{mca82, mca83, mca87, mca89, mca93, har93, mas98, pri01} 
indicates a linear motion widening by 5.7 mas y$^{-1}$.
The {\it Hipparcos} satellite found this companion at a 
separation of $0\farcs267 \pm 0\farcs003$ in 1993 and determined 
a magnitude difference between the Aa and Ab components 
of $\Delta H_p = 1.35 \pm 0.03$~mag \citep{per97}.

\placefigure{fig1}    

The bright central object, $\delta$~Ori~A (O9.5~II,
\citet{wal72}), is a single-lined spectroscopic binary
with a period of $5\fd7325$ \citep{har87}.
Orbital solutions exist which span the last century 
\citep{har04, jor14, cur14, hna20, luy39, pis50, mic52, nat71, sin82},
and these are summarized by \citet{har87} who present 
the most recent solution based upon 44 high dispersion 
spectra from {\it IUE}.
In addition, \citet{har87} reanalyzed all the published data 
and found convincing evidence of a regular advance in the longitude of 
periastron, giving an apsidal period of $225 \pm 27$~y.

Spectroscopic investigations to date have failed to identify 
clearly the secondary and tertiary (Hei~42~Ab) spectral 
components in the composite spectrum. 
The first detection of the secondary was claimed by 
\citet{luy39} who observed its orbital motion in weak extensions of  
\ion{He}{1} $\lambda 4471$ in 6 of their 140 photographic spectra 
(with a semiamplitude ratio of $K_2/K_1 = 2.6$).
\citet{gal76} reported evidence of the secondary star's contribution 
in the \ion{He}{1} $\lambda 5876$ profile, and found 
the antiphase motion amounting to $K_2/K_1 = 1.6 - 1.7$.   
\citet{gal76} estimated a spectral type of B1 for the secondary and a
magnitude difference of $\Delta m = 2.0-2.5$ mag in the optical region.

The central binary also exhibits partial eclipses, and 
\citet{koc81} have analyzed both their photometric observations 
and the available, historical light curves.   The analysis 
is made difficult by the presence of intrinsic light variability
in one or more of the components and the presence of third light. 
Nevertheless, \citet{koc81} were able to make a reasonable 
fit of the light curve by assuming a detached (but near 
Roche-filling) configuration.   Their best fit model 
led to a primary of $23~M_\sun$, $17~R_\sun$, and $T_{\rm eff} = 31100$~K,
a secondary of $9~M_\sun$, $10~R_\sun$, and $T_{\rm eff} = 25000$~K, 
an optical magnitude difference $\Delta m = 1.4$ mag, and 
an orbital inclination of about $i = 68^\circ$.
However, \citet{koc81} had to rely on Heintz's visual report of 
a third light magnitude difference of $\Delta m = 0.1$ mag, 
implying a much brighter Ab component than found by {\it Hipparcos}
($\Delta m = 1.35$ mag), and this suggests that their corrected eclipse 
depths are too deep. 

In this paper we present our own analysis of the 60 available
{\it IUE} high-dispersion SWP spectra of $\delta$~Ori~A 
plus a set of spectra of the \ion{He}{1} $\lambda 6678$ line. 
In the next section (\S2), we describe the {\it IUE} observations and our
cross-correlation methods for radial velocity measurement.
The optical spectra and radial velocity measurements are presented in \S3. 
A new double-lined orbit is given in \S4 along with a discussion 
of the apsidal motion and long term changes in systemic velocity. 
Next (\S5), we use Doppler tomography to separate the 
individual component spectra, which are then examined to obtain 
MK classifications, projected rotational velocities, and 
relative flux contributions.
We analyze the {\it Hipparcos} light curve in \S6, 
and discuss in \S7 the evolutionary implications of the resulting
masses and other parameters. 

\section{{\it IUE} Observations and Radial Velocities}

There are 60 high-dispersion, short wavelength prime (SWP) camera
spectra of $\delta$~Ori~A available in the {\it IUE} Final Archive
at the Space Telescope Science Institute's Multi-mission
Archive\footnote{Available, along with on-line documentation describing
{\it IUE}, its characteristics, and the available data products, at
http://archive.stsci.edu/iue.}.
These spectra have a signal-to-noise ratio (S/N) of approximately 10 per
pixel.
They were made with both the large and small apertures.
They were retrieved in NEWSIPS format \citep{gar97}.
A log of observations, including the SWP image number and 
the heliocentric time of mid-observation appears in Table~1.  

\placetable{tab1}     

Once we retrieved the spectra, they were processed using local
procedures written in 
IDL\footnote{IDL is a registered trademark of Research Systems, Inc.} 
that assemble the spectra into an array of dimensions of wavelength and time.
During this process, the spectra are smoothed from the nominal
30 ~km~s$^{-1}$ FWHM resolution of the raw spectra to 40 ~km~s$^{-1}$ FWHM.
The spectra are rectified using a set of relatively line-free zones,
and resampled using a uniform log $\lambda$ wavelength grid (in increments
equivalent to $10 {\rm ~km~s}^{-1}$).
The major interstellar lines are used to co-align the spectra on a common
velocity frame, and then these lines are replaced by straight-line segments.
Details are given in \citet{pen97}.

We then cross-cor\-rel\-at\-ed the spectra with 
the spectrum of a narrow-lined reference template star,
AE~Aurigae (HD~34078; O9.5~V; $V \sin i = 30 {\rm ~km~s}^{-1}$;
$V_r = +54.4 {\rm ~km~s}^{-1}$, \citet{gie86}).
The template was produced from an average of 6 SWP spectra of AE~Aur prepared
using the same methods that we used to produce the $\delta$~Ori matrix of
spectra.
The cross-correlation was performed over the velocity range
$-1000$ to $+1000 {\rm ~km~s}^{-1}$ relative to the reference star to produce
a matrix of cross-correlation functions (the composite ccfs)
of dimensions relative velocity and time.
The regions around the broad features of Ly$\alpha$, \ion{Si}{4} $\lambda 1400$,
and \ion{C}{4} $\lambda 1550$ were set to unit intensity (featureless continuum)
before cross-correlation to insure that the resulting ccfs have shapes 
similar to the rotationally broadened photospheric lines.  
Finally the ccfs were rectified by division of a straight line fit 
made of the extreme ends of the ccfs.  This step makes possible the direct 
comparison of ccfs made from spectra of varying quality (S/N ratio), but 
we caution that in this representation the ccf central depths may vary depending 
on spectrum quality. 

The final ccfs are presented in Figure~2 
as a function of absolute radial velocity and orbital phase. 
The orbital motion of the primary star (Aa1) is clearly evident
in the varying position of the central minimum and accompanying 
broad wings of the ccfs (termed ``pedestal'' by \citet{how97}). 
There is no obvious indication of a secondary (or a stationary 
tertiary) component, but we demonstrate below that the shapes of 
the ccfs are indeed influenced by the moving secondary component.  

\placefigure{fig2}    

We first attempted to measure radial velocities for the primary 
component by fitting a parabola through the 5
points (spanning $50 {\rm ~km~s}^{-1}$) surrounding the ccf minimum. 
This procedure proved unsatisfactory because the residuals 
from the fit were rather large (due to fitting only a small 
portion of the ccf) and because inspection of the quadrature 
phase ccfs showed asymmetrical extensions indicating the presence
of a partially blended, weak secondary component.   
Our solution to the fitting 
problem involved determining an empirical shape for the 
ccf component of the primary and a constrained Gaussian 
function for the secondary, and these components were then fit 
in a least-squares model to determine the positions (and 
hence radial velocity shifts) for each component.   
The following paragraphs describe the details of the 
fitting procedure.  Our final measurements are listed in 
Table~1, and are discussed below (\S4). 
 
The ccfs are dominated by the central core and broad wings 
of the primary component, and our first task was to isolate 
the functional shape of this component.
During orbital 
phases when the primary is significantly blue-shifted, 
the short wavelength half of the profile is mainly clear of 
the blending influence of the secondary (or a possible 
stationary third component) while the long wavelength side 
is free of blending at the opposite red-shifted phases. 
We therefore formed two versions of the primary component's
profile by using a shift-and-add algorithm and our preliminary
orbital solution to obtain the mean profile near the times of each quadrature. 
These two means were then averaged using window functions of the 
form $(1+e^x)^{-1}$ (with $x=\triangle V_r / 50~{\rm ~km~s}^{-1}$) 
for the blue-shifted mean and $(1+e^{-x})^{-1}$ for 
the red-shifted mean, to obtain a final primary reference ccf 
which varied smoothly through the line core (illustrated 
in Fig.~3).   

The next step was to define a functional shape for the 
secondary component in the ccf.
This was done by 
fitting the primary reference ccf to the best resolved quadrature 
phase ccfs over the range from the ccf minimum to the high velocity 
wing in the direction away from the secondary.
This fit was 
then subtracted away, and the residual ccf due to the secondary alone was 
fit with an unconstrained Gaussian profile.   
There were no obvious differences between 
the derived Gaussian widths or depths of the fits 
from the two quadratures, so we averaged all the Gaussian fit parameters
to arrive at a mean standard deviation width,
$\sigma = (89 \pm 8) {\rm ~km~s}^{-1}$,
and ratio of central depth to primary reference ccf depth of
$0.09 \pm 0.02$.
We adopted these parameters 
to fix the shape of the secondary ccf component. 

The final step was to fit the sum of these profiles to each individual
ccf. 
This was accomplished using a non-linear, least-squares fitting of the
primary and secondary functional curves based upon three parameters, the
primary and secondary radial velocities and a normalization factor for
overall ccf depth (to account for varying quality of the input
spectra).
An example of the detailed fitting procedure is illustrated in Figure~3
which shows the model primary and secondary functions and how their sum
gives excellent agreement with the observed ccf (for a phase near primary
velocity maximum).
This plot demonstrates the importance of defining the shape of the primary
component wings in order to place correctly the secondary
component.
The residuals formed after removal of only the primary component in these fits
are shown in Figure~4.
This plot shows that although the secondary component is weak, it is nevertheless
found in most of the ccfs, moves with the expected period, and is antiphase to
the primary's orbit.
The average ccf variance per pixel in the inner $\pm 200$~km~s$^{-1}$ of the velocity
range in Figure~4, the range that bounds the secondary's Doppler shifts, is 1.5\% (or
about one sixth of the height of the scale bar illustrated) based upon differences
in ccfs from multiple spectra obtained sequentially. 
This is $\approx 67\% $ of the the secondary's depth in the residual ccfs, 
so the secondary's ccf contribution is slightly greater than 
the noise level in individual pixels.
The final velocity measurements on the absolute scale are listed in Table~1, 
and the radial velocity curves are illustrated in Figure~5. 

\placefigure{fig3}    

\placefigure{fig4}    

\placefigure{fig5}    

These results indicate that the secondary's contribution to the composite
ccf is always weak and partially blended with the primary's ccf component. 
It is therefore important to review the limitations of our methods 
and the possible systematic errors that may be present in these difficult 
measurements.
The first shortcoming is that there probably exists 
an additional stationary component in the composite ccfs that is not 
included in our simple two-component model.
The grayscale portion of   
Figure~4, for example, indicates that there is a very weak, narrow, and
stationary 
feature in the residuals near a radial velocity of $+54 {\rm ~km~s}^{-1}$ 
which we suggest results from a correlation between unexcised interstellar 
lines in the spectrum of $\delta$~Ori with their counterparts in the 
AE~Aur photospheric spectrum (since this velocity corresponds to the 
radial velocity of AE~Aur).
It is also possible that there is a  
weak, stationary ccf component near the systemic velocity due to the 
tertiary (although there is no evidence of this in 
Fig.~4 nor from tomographic experiments described below).
Any such 
stationary feature would tend to pull the two-component fits back 
towards the stationary component, so that the velocity excursions 
would be underestimated in our measurements.
We did a numerical experiment
to estimate the size of this error by artificially removing from the 
composite ccfs a stationary component to represent a possible tertiary
contribution.
We assumed that the third component had a shape like the primary's 
reference ccf but was reduced to 10\% of the primary's central depth and 
centered at the systemic velocity.
We then repeated the measurement 
process by forming a primary reference profile from the quadrature phase
ccfs, fitting a Gaussian to the isolated secondary component, and
then fitting all the adjusted ccfs for the positions of primary 
and secondary.
These fits were generally less satisfactory and
we do not list the individual velocities here, but the orbital 
elements based upon this experiment are given in the fourth column
of Table~3 under the heading ``Adjusted UV ccfs''.
A comparison 
with our nominal set of elements (third column of Table~3) shows 
that the semiamplitudes may be underestimated (by $\approx$ 8\% for 
$K_1$ and 14\% for $K_2$ in this particular case) if such a third
component is present.  
This effect, if fully present in both components, could raise our
estimated masses by 43\% (\S6).

A second area of concern is the parameterization of the 
secondary component's width and depth.
Our choices for these 
parameters appear to be supported by observations of the optical 
line \ion{He}{1} $\lambda 6678$ (\S3), but the results are
sensitive to the values of these parameters.  
If the secondary profile is set to be wider and/or deeper, then 
the observed ccfs are fit with smaller radial velocity 
excursions.
For example, we made a test with a secondary Gaussian component 
with a mean standard deviation, $\sigma = 128 {\rm ~km~s}^{-1}$, and a 
ratio of central depth to primary reference ccf depth of $0.10$, 
and the resulting measurements led to a secondary semiamplitude of 
$K_2 = 112 {\rm ~km~s}^{-1}$ which is much smaller than our nominal
(preferred) value of $K_2 = 186 {\rm ~km~s}^{-1}$.
Finally, the reader should be aware of potential problems resulting
from our assumption that the component ratios and functional shapes
are constant with orbital phase.
At the conjunction phases for example, the partial 
eclipses will change the relative contributions of the primary and 
secondary components so that our fixed ccf depth ratio may lead to 
systematic errors in the velocity placement.
Furthermore, the photometric 
light curve indicates that the primary is close to Roche filling 
\citep{koc81}, and consequently we expect that the primary's 
photospheric lines will show subtle width variations, appearing 
slightly narrower (wider or asymmetric) at conjunctions (quadratures).   
However, a comparison of our primary reference ccf (which was 
formed from the quadrature ccfs) with conjunction phase ccfs 
indicates only minor differences so we doubt that our 
radial velocity measurements are badly in error at conjunctions. 

\section{Radial Velocities from \ion{He}{1} $\lambda6678$}

The binary was also observed in the optical by \citet{tha97a} in a 
program to search for incipient H$\alpha$ emission from colliding winds. 
\citet{tha97b} found that H$\alpha$ is sometimes distorted by 
residual emission effects, but \ion{He}{1} $\lambda6678$
is relatively free of emission problems most of the time. 
There are 20 spectra in this set which were obtained 
with the Kitt Peak National Observatory Coude Feed Telescope and 
the Mount Stromlo Observatory 74-inch Telescope \citep{tha97a},
and these spectra generally have better resolution 
($\lambda /\Delta\lambda = 14000 - 32000$) and S/N ratio 
($\approx 200 - 300$ pixel$^{-1}$) than the {\it IUE} ccfs discussed above.   
The \ion{He}{1} $\lambda6678$ profiles are illustrated 
as a function of orbital phase in Figure~6.   There appears 
to be a weak extension present at both quadratures that 
is probably the signature of the secondary (although part of 
the red wing depression is due to a weak \ion{He}{2} $\lambda 6683$
feature in the primary's spectrum).   

\placefigure{fig6}    

The \ion{He}{1} $\lambda6678$ profiles were fit in the same way 
as the UV ccfs by forming a mean quadrature profile 
for the primary and fitting the residual secondary component 
with a Gaussian with a mean standard deviation,
$\sigma = (86 \pm 20) {\rm ~km~s}^{-1}$
and ratio of central depth to primary depth of 
$0.10 \pm 0.02$ (both of these parameters are identical within errors 
with those found from the UV ccfs).  The fits were generally good, 
and we show in Figure~7 the residuals after the removal of the 
fitted primary component alone.   The motion of the secondary 
component is clearly seen in these residuals.  The radial velocities from 
the two-component fit are listed in Table~2 and plotted in Figure~8. 
Note that we zero weighted three sets of measurements in the orbital 
solutions below that corresponded to profiles which were visibly 
contaminated with the emission pattern seen in H$\alpha$. 

\placetable{tab2}     

\placefigure{fig7}    

\placefigure{fig8}    

\section{Orbital Elements}

We solved for the orbital elements using the non-linear, least-squares 
fitting program described by \citet{mor74}.
The fit progressed in several steps.
First, we fit the UV ccf radial velocities of the primary component 
by solving for all six orbital elements.
Since the radial velocity measurements of the secondary generally
have larger errors, we then 
fixed the values of the period, $P$, epoch of periastron, $T$, 
eccentricity, $e$, and longitude of periastron, $\omega$, to those 
from the primary's solution, and solved only for the secondary systemic
radial velocity, $\gamma$, and semiamplitude, $K$.
In Table~3, we compare the results
from the UV ccf velocities (column 3) with those of 
\citet{har87} (column 2).
This table also lists the 
elements resulting from a numerical experiment (\S2) which 
shows the influence of a potentially unaccounted for tertiary component
in our fitting of the UV ccfs (column 4).
We have relatively fewer radial velocities from 
fits of the \ion{He}{1} $\lambda6678$ profile, and so we fixed the 
period and eccentricity at the values from the UV ccf solution 
in fitting the \ion{He}{1} $\lambda6678$ data (column 5).
The computed radial velocity curves are plotted in Figures 2 and 4~--~8.  

\placetable{tab3}     

Our results (with the exception of $\omega$ and $\gamma$; see below) are 
generally in good agreement with the single-lined spectroscopic orbital
solution found by \citet{har87}, and our formal errors are similar to or 
smaller than theirs.  We give in Table~3 the mean sidereal period rather 
than the anomalistic period given by \citet{har87}. 
Our derived period agrees with theirs within errors, although we 
find somewhat better agreement with the photometric
period  of 5.732476~d adopted by \citet{koc81}. 

The largest errors are associated with the secondary semiamplitude, 
$K_2$, which unfortunately propagates into a large error in 
the quantity $m_1 \sin ^3 i$.  
The root-mean-square (rms) residuals are smaller 
for the secondary radial velocity measurements from the higher quality 
\ion{He}{1} $\lambda6678$ profiles.
Our value of $K_2$ is 
much smaller than first claimed by \citet{luy39} but is slightly 
larger than that found by \citet{gal76}.
Given the faintness and line blending problems associated with
the secondary's lines, 
this kind of disagreement is not surprising. 

The greatest difference between our results and prior work is 
the evident increase in longitude of periastron, $\omega_1$, 
due to apsidal motion.   
\citet{har87} performed a comprehensive non-linear, least squares
analysis of the entire body of 546 historical and new velocities 
using $\dot{\omega}_1$ as a free parameter and fixing the 
systemic velocity at $+20.3 {\rm ~km~s}^{-1}$.
They found an apsidal motion of $(1.60 \pm 0.20)$ $^\circ$~y$^{-1}$,
giving a period of $225 \pm 27$~y, in excellent agreement with
an earlier comprehensive result of $227 \pm 37$~y from \citet{mon80}. 
We have added our results from the expanded {\it IUE} ccf
and \ion{He}{1} $\lambda 6678$ solutions to the historical 
derivations of $\omega_1$ given in \citet{har87}, and 
the entire set is plotted as a function of time in Figure~9. 
The error weighted, linear fit is also shown which yields  
the same rate for the apsidal advance as found by \citet{har87},
$(1.60~\pm~0.03)~^\circ$~y$^{-1}$, giving an apsidal period of
$224.5 \pm 4.5$~y.  
Note that apsidal advance in $\delta$~Ori may result from 
both the tidal forces between the two components of the close binary
and the influence of the orbiting third body.

\placefigure{fig9}    

It is possible that the reflex motion of the central binary 
in its orbit about the center of mass of the binary-tertiary system
might produce measurable, long-term variations in the 
systemic velocity of the central binary.   
We show in Figure~10 estimates of the systemic velocity 
from solutions based on data spanning the last century \citep{har87} 
plus our own estimates from Table~3. 
There is a suggestion in this plot of a real variation
in which the systemic velocity reached a minimum around 1940 
and is currently increasing.
The {\it IUE} solution corresponds 
to the highest point in this graph, and we caution that 
the apparent discrepancy may be due in part to systematic 
differences between radial velocities for the lines in the 
UV and optical.
\citet{hut76} demonstrated that velocity 
progressions exist in hot, luminous stars between lines formed 
deep in the photosphere and those of lower excitation formed 
higher in the atmosphere where acceleration into the stellar 
wind is occurring.
Many of the strong optical lines 
(the Balmer series and \ion{He}{1} lines) probably reflect the 
outflow as a negative radial velocity offset compared to the 
higher excitation lines measured in the UV ccfs (formed at 
higher temperature deeper in the photosphere), and thus, the 
systemic velocity from the optical studies would be lower than 
that from our {\it IUE} solution.
We note that \citet{hut76}
found no measurable velocity -- excitation correlation in his optical
spectra of $\delta$~Ori, but a difference of $10 {\rm ~km~s}^{-1}$ 
between the radial velocities of the low and high excitation lines is
smaller than his threshold for detection. 

\placefigure{fig10}    

\section{Tomographic Reconstruction of the Individual Spectra}

We used an iterative Doppler tomography algorithm \citep{bag94} to reconstruct
the spectra of the individual component stars in the $\delta$ Orionis A system.
The algorithm assumes that each observed composite spectrum is a linear 
combination of spectral components with known radial velocity curves 
(Table 3) and flux ratios that are constant across the spectral range and
throughout the orbital cycle.  The latter condition is violated 
by the partial eclipses that occur at conjunctions (\S5), but since the eclipses 
are shallow and only a minority of our spectra were obtained during 
eclipse phases, flux constancy is a reasonable assumption. 
We estimated the UV flux ratio, $F_s/F_p(UV) = 0.095 \pm 0.021$, 
from the ratio of the areas of Gaussian fits of 
the UV ccf reference profiles.   Note that the actual error is 
probably larger than quoted since we have not 
accounted for changes in the ccf amplitude related to differences
in the match of the ccf template spectrum of AE~Aur to the 
slightly different spectral patterns of the primary and secondary. 
Likewise, we estimated the optical band flux ratio as 
$F_s/F_p(opt) = 0.097 \pm 0.027$ based upon the ratio of the
\ion{He}{1} $\lambda 6678$ reference component equivalent widths. 
Again, this estimation does not account for differences in equivalent 
width of \ion{He}{1} $\lambda 6678$ between the spectral types 
of the primary and secondary.   We took the {\it Hipparcos} 
magnitude difference for the tertiary to estimate its optical 
flux ratio as $F_t/(F_p+F_s)(opt) = 0.288 \pm 0.009$.  
Note that the tertiary is actually the second brightest star 
of the three in the visual region.  We do not 
know the UV flux contribution of the tertiary, so we produced 
UV reconstructions assuming both zero flux and the optical flux ratio. 

The reconstructed spectra in the vicinity of \ion{He}{1} $\lambda 6678$
are illustrated in Figure~11.  The primary and secondary components
both have a strong \ion{He}{1} $\lambda 6678$ feature, and 
\ion{He}{2} $\lambda 6683$ is also visible in the red wing of 
the primary's profile.   The tertiary, on the other hand, 
has a very shallow and broad \ion{He}{1} $\lambda 6678$ feature 
which is shifted redward by $27 {\rm ~km~s}^{-1}$ compared to the 
systemic velocity of the central binary.   The equivalent 
width of the profile in the tertiary reconstruction is comparable 
to those of the primary and secondary, and this suggests that 
the tertiary has an early B-type classification.   The extreme 
width of the reconstructed tertiary profile indicates that 
either it is a rapid rotator ($V\sin i \approx 300 {\rm ~km~s}^{-1}$) 
or that it is a spectroscopic binary with a radial velocity 
range similar to the line width in the reconstruction. 

\placefigure{fig11}    

A portion of reconstructed UV spectrum of the primary appears 
in Figure~12, and the S/N in the reconstructed spectrum is excellent 
thanks to the large number of spectra used in the reconstruction. 
\citet{pen97} developed a system of spectral classification 
in the UV based upon the equivalent widths of a set of 
prominent photospheric spectral lines that are sensitive to
temperature and luminosity.   We used this method with 
the reconstructed primary spectrum to derive a 
classification of O9.5~III  based on the two component reconstruction or 
B0~I based on the three component reconstruction (in which the 
primary has deeper lines) using the following criteria:  
the equivalent widths of 
\ion{Si}{3} $\lambda 1299$,
\ion{Fe}{5} $\lambda 1429$, 
\ion{He}{2} $\lambda 1640$, 
\ion{Fe}{4} $\lambda 1681$, 
\ion{Fe}{4} $\lambda 1765$, and the line ratios of
\ion{He}{2} $\lambda 1640$/\ion{Fe}{5} $\lambda 1429$ and
\ion{Fe}{4} $\lambda 1723$/\ion{Fe}{5} $\lambda 1429$. 
The luminosity class is set by the equivalent width of 
\ion{N}{4} $\lambda 1718$ which is used to assign a 
class of I, III, or V.   These UV-based criteria 
are fully consistent with the optical classification 
of O9.5~II \citep{wal72}.   

\placefigure{fig12}    

The depths of the lines in the reconstructed spectrum of 
the primary offer a means to check on our assumed flux 
contribution of the tertiary.
If we assigned too much flux 
to the tertiary, for example, then the primary's lines will 
appear too deep in the reconstructed spectrum. 
We can test this possibility through a direct comparison 
of the reconstructed spectrum with that of a single star 
of the same spectral classification and rotational broadening. 
We selected 19~Cep (HD~209975) for this purpose which 
has a similar classification (O9.5~Ib) but slightly 
narrower lines ($V\sin i = 90 {\rm ~km~s}^{-1}$; \citet{pen963}). 
We used the reference ccf widths to estimate the projected 
rotational velocities using the method of \citet{pen963} 
to arrive at $V\sin i = 157\pm 6 {\rm ~km~s}^{-1}$ and 
$138\pm 16 {\rm ~km~s}^{-1}$ for the primary and 
secondary, respectively.
Then we artificially broadened the spectrum of 19~Cep to match the
line broadening of the primary. 
The resulting spectrum of 19~Cep ({\it dotted line}) is 
compared to that of the primary in Figure~12 for the cases where the
tertiary contributes no UV flux ({\it thick solid line})
and where it contributes the fraction measured in the optical 
({\it thin solid line}).
Taken at face value, this comparison 
indicates that the tertiary contributes little if any UV flux. 

The reconstructed secondary spectrum is noisy because
of the star's relative faintness, and this reconstruction is 
also marred by the effects of incomplete removal of the 
echelle blaze function (``ripple'') in the reduction software (Fig.~13).
Furthermore, several features useful for spectral 
classification in B-type stars were excised during the 
process of interstellar line removal 
(e.g., \ion{Si}{2} $\lambda 1265$, \ion{C}{2} $\lambda\lambda 1334, 1336$).
Nevertheless, a number of prominent spectral features are 
discernible and indicative of an early-B spectral type. 
Since the classification scheme of \citet{pen97} is 
valid only for O-type stars, we relied instead on the 
classification criteria described by \citet{rou93} and
\citet{mas89}.   \citet{mas89} found that the equivalent 
width of \ion{Si}{3} $\lambda 1417$ is luminosity sensitive
in early-B stars, and the measured value in the secondary,
$\log W_\lambda (${\rm m\AA}$) = 2.49\pm 0.04$ indicates 
luminosity class III which is consistent with the observed
strength of other luminosity sensitive lines discussed 
by \citet{rou93}.   Based upon a comparison of the relative line 
strengths in the spectra of giants presented in \citet{rou93}, the 
spectral type falls in the range 
B0--B1 using \ion{C}{3} $\lambda 1247$, 
B1--B3 using \ion{Si}{3} $\lambda 1299$, 
B0.5--B1 using \ion{He}{2} $\lambda 1640$, 
B0--B0.5 using \ion{N}{4} $\lambda 1718$, and 
B0.5--B1 using \ion{N}{3} $\lambda\lambda 1748,1751$. 
Thus, the secondary has a UV spectrum consistent with 
a B0.5~III classification.   The secondary spectrum is 
compared in Figure~13 to the co-added spectrum of a similar star, 
$\epsilon$~Per (B0.5~IV-III, $V\sin i = 134 {\rm ~km~s}^{-1}$; 
\citet{tar95,gie99}).
The reconstructed secondary does exhibit P~Cygni type 
profiles of \ion{N}{5} $\lambda 1240$, \ion{Si}{4} $\lambda 1400$, 
and \ion{C}{4} $\lambda 1550$ indicative of an earlier or more luminous star, 
but we caution that the tomographic reconstructions are problematical
in the immediate vicinity of these lines which often lack 
obvious orbital motion \citep{gie95}.  Our B0.5~III classification
is consistent with the optical detections of the secondary's lines 
in \ion{He}{1} $\lambda 4471$ \citep{luy39},  
\ion{He}{1} $\lambda 5876$ \citep{gal76, ful90}, and in our 
tomographic reconstruction of \ion{He}{1} $\lambda 6678$ (Fig.~11). 

\placefigure{fig13}    

In spite of the fact that our algorithm has successfully detected
tertiary components in the cases of $\delta$ Cir \citep{pen01} and
55 UMa \citep{liu97}, the reconstruction of the UV spectrum of the
tertiary (made assuming 
it was stationary over the time span of the {\it IUE} observations)
is essentially featureless, so that its UV spectral features 
remain undetected.
We show in Figure~14 a portion of this spectrum 
plotted with that of the rapid rotator, $\zeta$~Oph 
(O9.5~V, $V\sin i = 400 {\rm ~km~s}^{-1}$; \citet{rei93}).  
This comparison spectrum demonstrates how rapid rotation causes 
most features to become blended into a pseudo-continuum, so that 
a rapidly rotating tertiary would be difficult to detect in the 
reconstruction.
However, this comparison also shows that 
certain strong features (such as \ion{N}{4} $\lambda 1718$) 
should still be relatively deep in the UV spectra of rapid rotators, 
and their absence in the tertiary spectrum reinforces the impression 
derived from the depths of the primary's lines that the 
tertiary contributes only a minor fraction of the UV flux
(certainly less than it does in the optical).
Such a low UV flux contribution is puzzling given the relative brightness
of the tertiary in the optical and the indication from the 
\ion{He}{1} $\lambda 6678$ reconstruction that it is a B-type star.
On the other hand, the lack of features in the tertiary spectrum 
suggests that the radial velocities derived from our two component 
fits of the UV ccfs should be relatively free of systematic 
errors due to line blending with the tertiary spectrum (so that 
the trial orbital elements in column 4 of Table~3 based on 
a tertiary with lines 10\% as deep as the primary's represents
an extreme and unlikely case). 

\placefigure{fig14}    

\section{Light Curve and Masses}

Photometry from the {\it Hipparcos} Satellite \citep{per97} 
provides a contemporary light curve for the central binary (Fig.~15), 
and although this light curve suffers from gaps in coverage and  
additional intrinsic variability \citep{koc81}, we will compare it to model 
orbital variations to obtain preliminary limits on the system inclination, $i$ 
(and hence the masses).
We used the light curve synthesis code GENSYN \citep{moc72} 
to produce model $V$-band differential light curves (almost identical to
differential {\it Hipparcos} magnitudes for hot stars).
The code was written for
binaries with circular orbits, and we used multiple runs to synthesize
the varying orbital separation in an elliptical orbit \citep{pen99}.
Our approach was to make a constrained fit using as much data as possible
from our spectroscopic results.   
The orbital parameters were taken from the UV ccf solutions
(using an interpolated longitude of periastron, 
$\omega = 128^\circ$, for the average time of observation, 
BY 1991.2; see Fig.~9).
The stellar temperatures and gravities were taken from 
\citet{voe89} for the primary ($T_{\rm eff} = 33,000$~K and 
$\log g = 3.4$), and for the secondary we used estimates from \citet{tar95} for 
a comparable star, $\epsilon$~Per ($T_{\rm eff} = 27,600$~K and $\log g = 3.8$).
We then estimated the physical fluxes and limb darkening coefficients from
tables in \citet{kur94} and \citet{wad85}, respectively.
The secondary's polar radius was set at $0.37~R_p$ (where $R_p$ is 
the assumed primary's polar radius) based upon our estimated magnitude 
difference, $\Delta V = 2.5$ (\S5), and the calculated surface fluxes. 
Finally, we set the rotational periods of the stars based upon the 
assumed values of $R_p$ and $i$ and the measured projected rotational velocities (\S5). 
Each trial run of GENSYN was defined by two model parameters, $R_p$ and $i$, 
and all the resulting light curves were corrected for the presence of third light 
using the {\it Hipparcos} flux ratio, $F_t/(F_s + F_p)= 0.29$.

\placefigure{fig15}    

The {\it Hipparcos} light curve is illustrated in Figure~15 as a function 
of photometric phase based upon the predicted time of minimum light 
according to the ephemeris from the UV ccfs for the primary (Table~3). 
We also show two model light curves which provide only marginally 
satisfactory fits but which correspond to the probable limits 
of our fitting parameters.   The solid line is the model light curve 
for $R_p/R_\odot = 13$ and $i=67^\circ$ which represents the largest 
size primary consistent with the fitting scheme.  The primary in this model
has a polar radius only 1\% smaller than the critical Roche radius 
at periastron.   The Roche radius can be somewhat larger than $13~R_\odot$ in 
models with lower inclination (since the system dimensions are 
set by $a\sin i$), but Roche-filling models with lower inclination 
cannot produce eclipses as deep as those observed, so $i=67^\circ$
represents a reliable lower limit for the inclination. 
This is fortuitously close to the inclination adopted by \citet{koc81}, 
$i=68^\circ$, since they adopted very different values for the 
mass ratio and third light correction.   This near Roche-filling 
model is characterized by large ellipsoidal variations that shape the 
light curve outside of the eclipses.   These appear to match the 
observations better in the photometric phase range 0.6 -- 0.9 than 
in the range 0.2 -- 0.4.  In fact, the observations in the 0.2 -- 0.4 range 
are better fit by a smaller, less tidally distorted model primary, 
and the dashed line indicates such a model for $R_p/R_\odot = 9$ and $i=77^\circ$.
This smaller primary model does a better job of matching the relative
depths of the eclipses, but the duration of the eclipses may be too short 
especially when compared to the light curve observations of \citet{koc81}. 
Thus, the inclination range $67^\circ - 77^\circ$ appears to span the 
acceptable limits of plausible fits.   This range corresponds to 
primary and secondary masses of $9.4 - 11.2~M_\odot$ and 
$4.8 - 5.6~M_\odot$, respectively (based upon the UV ccf orbital solutions). 

\section{Discussion}

The {\it IUE} ccfs and \ion{He}{1} $\lambda 6678$ line profiles have 
provided us with the first clear and consistent measurements of the 
orbital motion of the secondary star.
However, the secondary's resulting semiamplitude is surprisingly small,
and consequently, our results indicate a binary with a smaller semimajor
axis (implying lower masses) than expected.
In particular, the mass of the primary is a factor of 3 to 4 below that
predicted for a normal single star with its classification and stellar
wind properties \citep{voe89,how89}. 

Such low mass results are not unprecedented.
For additional examples of orbital masses that are significantly lower
($\approx$ 40\%) than theoretical expectations, see the cases of \mbox{LZ Cep}
\citep{har98} and \mbox{$\delta$~Cir} \citep{pen01}.
In addition, \citet{her99, her00} use model atmospheres to find masses
for single O stars, and in some cases, they also find masses that are
significantly lower than those expected from evolutionary tracks.

To attain a primary mass of $30M_\odot$ would require a 51\% upward
revision in the secondary's semiamplitude, to $K_2 = 281 {\rm ~km~s}^{-1}$ 
(for $i=67^\circ$ and $K_1=94.9 {\rm ~km~s}^{-1}$), placing the secondary's
radial velocity range between $-230$ and $+330 {\rm ~km~s}^{-1}$.
This would appear to be ruled out by the observed quadrature line
profiles (see Figs.~4 and 7).
Our derived value of $K_2$ (and the small masses implied) clearly
requires additional verification (through high S/N and high dispersion
spectroscopy of metal lines unaffected by emission, for example), but
taken at face value, our results suggest that the current state of the
primary is the result of extensive mass loss caused by an interaction
with the secondary.

The absolute magnitudes of the components provide a reference point to
assess the apparent \mbox{mass~-~luminosity} discrepancy of the
components. 
The three bright Belt stars in Orion, $\delta$~Ori, $\epsilon$~Ori (B0~Ia), 
and $\zeta$~Ori (O9.7~Ib), form the high luminosity end of the Ori~OB1b
subgroup \citep{bro94,dez99} which has an estimated age between 1.7~My
\citep{bro94} and 7~My \citep{bla91}.  
The distance to Ori~OB1b is estimated to be $360\pm 70$~pc based on
photometry \citep{bro94} or $473\pm 33$~pc based on {\it Hipparcos}
parallax measurements \citep{dez99}.
Outside of eclipse, the apparent magnitude of $\delta$~Ori is $m_V=2.19$
(see Fig.~15 and the $H_p$ to $V$ transformation in \citet{hrm98}),
which for an extinction of $A_V=0.16$~mag \citep{bro96}, gives the
absolute magnitude of the triple as $M_V=-5.8$ at $d=360$~pc or $M_V=-6.3$
at $d=473$~pc.
The absolute magnitude is also calculated in the light curve model (\S6)
based on the assumed sizes and $V$-band fluxes of the
stars.
The predicted absolute magnitudes for the triple are $M_V=-5.8$ for the
$R_p/R_\odot = 13$ model (near Roche-filling) and $M_V=-4.9$ for the
$R_p/R_\odot = 9$ model.
Thus, we can obtain consistency between these estimates if we adopt the
closer distance for Ori~OB1b and the larger radius primary model for the
binary.
A distance of 360~pc corresponds to a parallax of 2.78 mas, which falls
within one standard deviation of the Hipparcos parallax for $\delta$~Ori,
$3.56 \pm 0.83$~mas.
The component absolute magnitudes are then $M_V=-5.4$ for the primary, 
$-2.9$ for the secondary, and $-4.2$ for the
tertiary. 
The adopted temperatures and radii result in $(\log T_{\rm eff}, \log L/L_\odot)$ pairs of
(4.52, 5.26) and (4.44, 4.08) for the primary and secondary, respectively.

The non-rotating single star evolutionary tracks of \citet{sch92} that pass
through the observed $(\log T_{\rm eff}, \log L/L_\odot)$ points correspond
to current masses of about $28 M_\odot$ and $12 M_\odot$ (and ages of 4 and
3~My) for the primary and secondary, respectively.  
Evolutionary models that include rotation \citep{mey00,heg00} indicate that 
rapid rotators with temperatures and luminosities comparable to 
the components of $\delta$~Ori~Aa may appear as much as 0.1 dex 
more luminous than non-rotating stars of the same mass.  
By applying an artificial reduction in luminosity of this
amount, the resulting evolutionary masses would be approximately 
$M/M_\odot = 26$ and 12 for the primary and secondary, respectively. 
All of these mass estimates are much larger than those found for our
$R_p/R_\odot = 13$ light curve model, where we obtained maximum masses of
$M/M_\odot = 11.2$ and 5.6 for the primary and secondary, respectively.
Thus, both stars are extremely overluminous for their orbitally determined
masses.

We rule out stellar wind losses as the main mass loss mechanism, because the
amount of mass to be removed from the primary, $\approx 17 M_\sun$, is much larger
than the wind can carry in the time available 
(the current mass loss rate is estimated to be 
$\log \dot{M} [M_\odot ~{\rm y}^{-1}] = -6.3$; \citet{bie89}).
The winds are even weaker for stars in the secondary's mass
range.

We believe that we can accommodate these mass losses within the framework of
current models of Roche lobe overflow (RLOF).  
The relative youth of the system and the Ori~OB1b association suggests
that binary mass loss occurred while the stars were still in the 
core hydrogen burning (CHB) stage (Case~A mass transfer). 
If either star had begun extensive H shell burning or core He burning, we
would probably find an enrichment in the products of nuclear burning in
one or both stars due to mass transfer or extensive mixing, but in fact,
the abundances of both He \citep{voe89} and N \citep{wal76} in the 
primary appear to be solar-like.
There is no compelling evidence of the strong H$\alpha$ emission associated
with RLOF gas streams at the present time \citep{tha97b}, and so, we tentatively
conclude that $\delta$~Ori~Aa is a post-RLOF object that has recently completed
Case~A RLOF.

\citet{wel01} present a series of models of Case~A 
evolution based on the assumption of conservative mass transfer, 
and their results shed some light on the possible evolutionary status 
of $\delta$~Ori~Aa.  RLOF begins during the slow expansion of the 
donor during CHB, and continues until the mass ratio is reversed and 
the system separation grows.   The mass donor may end up 
as a very luminous and hot star at the conclusion of RLOF even though it 
has lost most of its mass \citep{van98,wel01}, and the mass gainer will 
then appear as the more luminous star (and perhaps overluminous for its 
mass due to accretion induced mixing; \citet{van98}). 
If this scenario is applied to $\delta$~Ori~Aa, then we must identify
the large, O9.5~II primary as the mass gainer and the 
smaller, B0.5~III secondary as the mass donor (the originally more massive star).  
The main difficulty with this explanation is the extreme overluminosity 
of the primary star (the presumed mass gainer).   We showed above that 
$11~M_\odot$ primary is radiating a luminosity comparable to that 
expected for a star more than twice as massive, and it remains to be
seen if processes like accretion induced mixing could account for such 
a large overluminosity.  

An alternative and more speculative explanation can be constructed 
based upon non-conservative RLOF.   If we assume that the high luminosity 
of the primary indicates it began life as a $\approx 28~M_\odot$ star, 
then we must conclude that most of its mass loss left the system entirely 
since the secondary is too small to have accreted any significant amount of mass. 
In this scenario the system started with a relatively large mass ratio, $M_p / M_s$, 
and models suggest that such systems may suffer mergers in a common envelope 
phase \citep{van98,wel01}.   It is possible that the $\delta$~Ori~Aa system 
experienced large scale mass loss during a common envelope phase, and the decrease 
in system mass caused an increase in binary separation, thereby avoiding a merger.  
The unusual infrared excess of $\delta$~Ori \citep{run96} may be a relic of this
substantial mass loss, and it would be worthwhile to search for any other evidence 
of large quantities of ejected gas. 

Finally we return to the mystery of the tertiary star.  
The tomographic reconstruction of the \ion{He}{1} $\lambda 6678$
profiles produced a very broad and shallow feature for the tertiary
spectrum (Fig.~11).
This result should be treated with some caution, since this feature
is subject to distortion by weak emission, but it would appear to
indicate that tertiary is a B-type star.
The width of the line suggests that it is either a rapid rotator or
a spectroscopic binary (averaged at many velocity displacements to
yield a net broadened profile in the reconstruction).
According to the absolute magnitude calibration of \citet{sch95}, a
main sequence star with the observed absolute magnitude of $-4.2$ has
a mass of $\approx$ $27 M_\odot$ (corresponding to an O8.5~V
classification; \citet{how89}).
On the other hand, if the tertiary were composed of two identical stars
of absolute magnitude $-3.5$, the stars would each have a mass of
$\approx$ $19 M_\odot$ (type B0.5~V).
In either case, we would expect the tertiary to produce a significant
UV flux contribution, comparable to or larger than the secondary's flux.
Our failure to detect the tertiary's UV spectral signature remains
unexplained.
The tertiary is probably gravitationally bound to the central binary
since otherwise its angular motion would have made it a striking double
to visual observers in the early centuries of telescopic observation.  
Consequently, we encourage continued spectroscopic and high resolution
observations to find the reflex orbital motion hinted at in Figure~10.   

\acknowledgments

Support for this work was provided in part by NASA through a 
grant from the Space Telescope Science Institute, which is 
operated by the Association of Universities for Research in 
Astronomy, Incorporated, under NASA contract NAS5-26555.
Institutional support was provided from the GSU College of Arts and Sciences 
and from the Research Program Enhancement fund of the Board of Regents of the
University System of Georgia administered through the GSU office of the
Vice President for Research and Sponsored Programs.
We gratefully acknowledge all this support.
This research made use of data obtained from 
the Multimission Archive at the Space Telescope Science Institute (MAST),
NASA's Astrophysics Data System Abstract Service,
the SIMBAD Astronomical Database of the Centre de Donn\'{e}es astronomiques de
Strasbourg, the Washington Double Star Catalog
and the Third Catalog of Interferometric Measurements of Binary Stars,
both maintained at the U.S. Naval Observatory, and the National
Institute of Standards and Technology's Atomic Spectra Database, Version 2.0
\footnote{Available at http://physics.nist.gov/PhysRefData/contents.html}.
We would like to thank an anonymous referee for a careful reading of the paper
and some very helpful suggestions.
We thank Dr.\ Anthony G.\ A.\ Brown for helpful comments.
Finally, we made use of an eclipsing binary simulator, Nightfall, written by
Dr.\ Rainer Wichmann, and freely available via the Internet.




\clearpage

\begin{deluxetable}{ccrrrrc}
\tablewidth{0pc}
\tablecaption{Journal of {\it IUE} Observations}
\tablehead{ \colhead{Date} & \colhead{Orbital} & \colhead{$V_1$} & \colhead{$(O-C)_1$}
& \colhead{$V_2$} & \colhead{$(O-C)_2$} & \colhead{{\it IUE}}  \\
\colhead{(HJD-2400000)} & \colhead{Phase} & \colhead{(km~s$^{-1}$)} & \colhead{(km~s$^{-1}$)}
& \colhead{(km~s$^{-1}$)} & \colhead{(km~s$^{-1}$)} & \colhead{Image ID} }

\startdata
43753.473 & 0.173 &  -65.0 &    3.9 &  174.6 &  -51.0 &   SWP02436  \\
43755.821 & 0.583 &   94.8 &   -1.2 & -158.5 &  -61.1 &   SWP02477  \\
43833.823 & 0.190 &  -64.8 &    3.2 &  155.7 &  -68.1 &   SWP03401  \\
43898.073 & 0.398 &    2.8 &   -6.2 &  161.3 &   88.3 &   SWP04018  \\
44126.004 & 0.159 &  -71.2 &   -2.4 &  129.3 &  -96.1 &   SWP06451  \\
44146.137 & 0.671 &  113.5 &   -4.6 & -163.0 &  -22.4 &   SWP06677  \\
44146.156 & 0.674 &  110.5 &   -8.0 & -164.2 &  -22.8 &   SWP06678  \\
44146.840 & 0.794 &  104.9 &   -4.1 & -165.6 &  -42.7 &   SWP06682  \\
44146.859 & 0.797 &  103.8 &   -4.2 & -145.9 &  -25.0 &   SWP06683  \\
44146.878 & 0.800 &   99.8 &   -7.2 & -170.6 &  -51.7 &   SWP06684  \\
44146.896 & 0.803 &  102.7 &   -3.3 & -155.3 &  -38.4 &   SWP06685  \\
44499.006 & 0.227 &  -60.0 &    2.1 &  156.3 &  -55.9 &   SWP10153  \\
44570.950 & 0.777 &  120.5 &    7.1 & -140.3 &   -8.8 &   SWP10690  \\
44631.629 & 0.362 &  -10.3 &   -1.1 &  163.2 &   54.5 &   SWP11163  \\
44631.693 & 0.373 &   -7.7 &   -4.1 &  149.0 &   51.4 &   SWP11166  \\
44631.711 & 0.377 &   -4.0 &   -2.0 &  148.6 &   54.1 &   SWP11167  \\
44652.439 & 0.992 &   -4.8 &   -0.6 &  142.5 &   43.7 &   SWP13320  \\
44652.504 & 0.004 &  -12.1 &   -0.6 &  134.3 &   21.1 &   SWP13323  \\
44676.498 & 0.189 &  -72.1 &   -4.0 &  167.0 &  -56.9 &   SWP13471  \\
44678.501 & 0.539 &   77.9 &   -0.8 & -183.9 & -120.4 &   SWP13491  \\
44679.183 & 0.658 &  116.6 &    0.6 & -161.7 &  -25.1 &   SWP13499  \\
45987.070 & 0.811 &  105.2 &    1.7 & -163.6 &  -51.5 &   SWP24169  \\
45991.784 & 0.633 &  110.4 &   -0.5 &  -51.6 &   75.0 &   SWP24197  \\
45991.947 & 0.661 &  114.5 &   -2.1 & -103.6 &   34.2 &   SWP24202  \\
45992.811 & 0.812 &  105.8 &    2.9 &  -35.1 &   75.9 &   SWP24208  \\
45993.041 & 0.852 &   90.5 &    5.1 & -104.0 &  -27.3 &   SWP24214  \\
45993.139 & 0.869 &   84.5 &    8.2 & -110.0 &  -51.1 &   SWP24217  \\
45993.831 & 0.990 &   -7.2 &   -4.6 &  172.4 &   76.6 &   SWP24230  \\
45994.086 & 0.035 &  -29.6 &    0.4 &  184.2 &   34.8 &   SWP24239  \\
45994.168 & 0.049 &  -38.7 &   -1.0 &  146.1 &  -18.4 &   SWP24241  \\
46075.734 & 0.278 &  -46.3 &    0.6 &  160.1 &  -22.4 &   SWP24878  \\
46076.521 & 0.415 &   14.3 &   -3.5 &  108.7 &   52.9 &   SWP24884  \\
46505.425 & 0.234 &  -58.0 &    2.3 &  168.3 &  -40.5 &   SWP27920  \\
46507.309 & 0.563 &   87.4 &   -1.2 &  -66.3 &   16.6 &   SWP27931  \\
46510.339 & 0.092 &  -52.9 &    3.2 &  188.1 &  -12.4 &   SWP27963  \\
46511.414 & 0.279 &  -42.1 &    4.1 &  200.4 &   19.3 &   SWP27975  \\
46511.551 & 0.303 &  -34.7 &    2.1 &  164.8 &    2.2 &   SWP27979  \\
46511.682 & 0.326 &  -28.9 &   -2.2 &  127.3 &  -15.5 &   SWP27982  \\
46511.821 & 0.350 &  -18.1 &   -3.0 &  121.8 &    1.5 &   SWP27985  \\
46512.345 & 0.442 &   39.0 &    7.3 &  -11.2 &  -39.7 &   SWP27990  \\
46512.471 & 0.464 &   43.5 &    0.5 &   54.4 &   48.0 &   SWP27994  \\
46512.609 & 0.488 &   61.8 &    6.8 &  -69.7 &  -52.6 &   SWP27998  \\
46512.743 & 0.511 &   71.0 &    4.7 &  -77.1 &  -38.0 &   SWP28001  \\
46515.438 & 0.981 &   -0.7 &   -4.0 &  172.2 &   88.0 &   SWP28028  \\
46866.743 & 0.264 &  -50.3 &    1.3 &  162.5 &  -29.1 &   SWP30490  \\
46867.370 & 0.374 &   -5.7 &   -2.2 &  186.5 &   89.0 &   SWP30496  \\
46867.761 & 0.442 &   40.4 &    8.6 &  -39.8 &  -68.1 &   SWP30506  \\
46868.314 & 0.538 &   82.4 &    4.0 &   -4.5 &   58.4 &   SWP30513  \\
46869.480 & 0.742 &  127.1 &    7.7 &  -72.3 &   70.9 &   SWP30524  \\
47044.515 & 0.276 &  -45.6 &    2.0 &  188.6 &    4.7 &   SWP31727  \\
47045.147 & 0.386 &    1.5 &   -1.1 &  132.8 &   47.3 &   SWP31745  \\
47045.640 & 0.472 &   47.2 &    0.1 &  -34.3 &  -32.7 &   SWP31760  \\
47045.808 & 0.501 &   66.3 &    4.8 &  -20.1 &    9.7 &   SWP31765  \\
47046.176 & 0.565 &   87.7 &   -1.7 & -116.8 &  -32.3 &   SWP31774  \\
47046.570 & 0.634 &  109.2 &   -2.0 & -126.8 &    0.3 &   SWP31783  \\
47046.868 & 0.686 &  117.7 &   -2.0 & -135.8 &    8.0 &   SWP31791  \\
47047.174 & 0.739 &  121.9 &    2.3 & -116.7 &   26.9 &   SWP31801  \\
48546.812 & 0.342 &  -23.5 &   -4.3 &  168.8 &   40.6 &   SWP42747  \\
48546.831 & 0.345 &  -21.0 &   -3.4 &  157.3 &   32.3 &   SWP42748  \\
48546.851 & 0.349 &  -20.2 &   -4.3 &  173.7 &   52.0 &   SWP42749 
\enddata
\end{deluxetable}


\clearpage

\begin{deluxetable}{ccrrrrc}
\tablewidth{0pc}
\tablecaption{Journal of Optical Observations}
\tablehead{ \colhead{Date} & \colhead{Orbital} & \colhead{$V_1$} & \colhead{$(O-C)_1$}
& \colhead{$V_2$} & \colhead{$(O-C)_2$} & \colhead{}  \\
\colhead{(HJD-2400000)} & \colhead{Phase} & \colhead{(km~s$^{-1}$)} & \colhead{(km~s$^{-1}$)}
& \colhead{(km~s$^{-1}$)} & \colhead{(km~s$^{-1}$)} & \colhead{Source} }

\startdata
49612.996 & 0.156 &  -44.8 &   -1.8 &  147.2 &    9.0 & KPNO 1994  \\
49612.997 & 0.156 &  -45.8 &   -2.9 &  148.2 &   10.2 & KPNO 1994  \\
49613.997 & 0.330 &   60.1 &   -6.3 &  -81.6 &  -27.3 & KPNO 1994  \\
49615.866 & 0.656 &   91.1 &   -3.8 & -114.0 &   -9.6 & KPNO 1994  \\
49615.867 & 0.657 &   92.0 &   -2.8 & -104.8 &   -0.6 & KPNO 1994  \\
49617.922 & 0.015 &  -80.5 &    8.8 &  235.2 &   15.5 & KPNO 1994\tablenotemark{a}  \\
50150.888 & 0.988 &  -58.5 &   28.1 &  220.0 &    5.0 & MSO 1996\tablenotemark{a}   \\
50151.885 & 0.162 &  -39.0 &    0.4 &  133.9 &    1.9 & MSO 1996   \\
50152.879 & 0.335 &   60.9 &   -8.0 &  -89.5 &  -30.9 & MSO 1996   \\
50153.883 & 0.510 &  117.1 &   -4.3 & -134.8 &   16.2 & MSO 1996   \\
50154.876 & 0.683 &   80.0 &   -3.0 & -100.5 &  -17.0 & MSO 1996   \\
50529.892 & 0.103 &  -82.5 &  -12.1 &  161.7 &  -24.8 & MSO 1997   \\
50530.874 & 0.274 &   40.5 &    7.3 &    1.7 &   -2.5 & MSO 1997   \\
50531.860 & 0.446 &  112.9 &    0.7 & -115.6 &   19.2 & MSO 1997   \\
50532.863 & 0.621 &  119.2 &   11.9 & -102.8 &   23.5 & MSO 1997   \\
50533.862 & 0.795 &   19.7 &    3.7 &   30.0 &   -4.4 & MSO 1997   \\
50534.882 & 0.973 &  -69.6 &   13.9 &  158.0 &  -51.6 & MSO 1997\tablenotemark{a} \\
50535.860 & 0.144 &  -36.4 &   13.5 &  181.5 &   31.1 & MSO 1997   \\
50536.864 & 0.319 &   69.4 &    9.4 &  -40.0 &    3.1 & MSO 1997   \\
50537.861 & 0.493 &  118.2 &   -1.9 & -145.7 &    3.0 & MSO 1997   \\
\enddata
\tablenotetext{a}{Assigned zero weight in the solution.}
\end{deluxetable}

\clearpage

\begin{deluxetable}{lcccc}
\tablewidth{0pc}
\tablecaption{Orbital Elements}
\tablehead{
\colhead{Element} &\colhead{\citet{har87}}  
&\colhead{UV ccfs}
&\colhead{Adjusted UV ccfs}
&\colhead{\ion{He}{1} $\lambda 6678$}
}
\startdata
$P$ (days)                   & 5.732424 (50)           & 5.732503 (26)         & 5.732508 (28)          & 5.732503\tablenotemark{a}  \\
$T$ (HJD -- 2,400,000)       & 30802.02 (9)            & 46865.23 (10)         & 46865.19 (11)          & 50535.0 (4)   \\
$e$                          & 0.087 (9)               & 0.075 (6)             & 0.076 (6)              & 0.075\tablenotemark{a}      \\
$\omega$ ($^\circ$)          & 50 (6)                  & 112 (6)               & 110 (7)                & 173 (23)     \\
$K_1$ (km s$^{-1}$)          & 97.9 (9)                & 94.9 (6)              & 102.5 (10)             & 105 (4)       \\
$\gamma_1$ (km s$^{-1}$)     & 20.3\tablenotemark{b}   & 28.6 (4)              & 27.5 (7)               & 24 (3)       \\
rms$_1$ (km s$^{-1}$)        & \nodata                 & 4.2                   & 4.9                    & 7.6            \\
$K_2$ (km s$^{-1}$)          & \nodata                 & 186 (9)               & 212 (11)               & 186 (7)     \\
$\gamma_2$ (km s$^{-1}$)     & \nodata                 & 35 (7)                & 36 (8)                 & 20 (4)      \\
rms$_2$ (km s$^{-1}$)        & \nodata                 & 51                    & 59                     & 18             \\
$m_1 \sin^3 i$ ($M_\sun$)    & \nodata                 & 8.7 (14)              & 12.4 (20)              & 9.3 (10)      \\
$m_2 \sin^3 i$ ($M_\sun$)    & \nodata                 & 4.4 (3)               & 6.0 (5)                & 5.3 (7)      \\
$a \sin i$ ($R_\sun$)        & \nodata                 & 31.7 (11)             & 35.5 (13)              & 32.9 (9)      \\
\enddata
\tablenotetext{a}{Fixed parameter.}
\tablenotetext{b}{\citet{har87} fixed $\gamma_1$ at 20.3 km~s$^{-1}$.}
\tablecomments{The parenthetic numbers are the standard errors in the last digit quoted (decimals omitted).}
\end{deluxetable}


\clearpage

\begin{figure}
\caption{
The components of the $\delta$~Ori system.
A, B, and C constitute the visual triple star, while 
Aa and Ab are the components of the speckle binary.
Aa1 and Aa2 are the primary and secondary stars, respectively, in the
eclipsing spectroscopic binary.
\label{fig1}}
\end{figure}

\begin{figure}
\caption{
The 60 cross-correlation functions made from the {\it IUE} spectra of
$\delta$~Ori~A, arranged in phase order according the `UV ccfs'
orbital solution given in Table 3 and plotted against heliocentric radial velocity.  
The ccfs are plotted in the upper portion with
their continua aligned with the orbital phase
of observation. The lower portion is a grayscale representation
of the ccfs (linearly interpolated in phase) with the calculated
radial velocity curves over plotted ({\it white lines}); arrows 
on the right hand side indicate the phases of observation. 
The first and last 20\% of the orbit have been reproduced 
at the bottom and top of the image, respectively, to improve 
the sense of phase continuity. 
\label{fig2}}
\end{figure}

\begin{figure}
\caption{
The primary and secondary ccf models (\S2) fitted to the cross-correlation
function of spectrum SWP31801.
The plus symbols show the cross-correlation
function of the composite spectrum which is compared to 
the model primary component ({\it long-dashed line}), 
model secondary component ({\it short-dashed line}), 
and their sum ({\it solid line}).
\label{fig3}}
\end{figure}

\begin{figure}
\caption{
The cross-correlation function residuals (in the same format as Fig.~2) 
after removal of the fitted primary component.  
The orbital motion of the secondary is now evident. 
\label{fig4}}
\end{figure}

\begin{figure}
\caption{
The combined {\it IUE} radial velocity curves of both stars.
The filled circles show the primary star radial velocities while
the open circles represent the secondary star radial velocities.
The first and last 20\% of the orbit are 
reproduced at the right and left 
to improve the sense of phase continuity.
The solid lines show the curves from the orbital solutions.
\label{fig5}}
\end{figure}

\begin{figure}
\caption{
The 20 profiles of \ion{He}{1} $\lambda 6678$, arranged in phase order 
according the \ion{He}{1} $\lambda 6678$ orbital solution in Table 3
(in the same format as Fig.~2).
Note that significant phase gaps in this data set cause some distortion
of the grayscale image (formed by linear interpolation in phase) here
and in Fig.~7.
\label{fig6}}
\end{figure}

\begin{figure}
\caption{
The residuals from the \ion{He}{1} $\lambda 6678$ profiles 
after removal of the fitted primary star's
profile (in the same format as Fig.~2). 
\label{fig7}}
\end{figure}

\begin{figure}
\caption{
The combined \ion{He}{1} $\lambda 6678$ 
radial velocity curves in the same format as Fig.~5.
The three observations affected by emission filling 
(made near phase 0.0) were given zero weight in the 
orbital solution.  
\label{fig8}}
\end{figure}

\begin{figure}
\caption{
The variation in the longitude of periastron,
$\omega_1$, over the last century.
The data are from \citet{har87} plus our results from
Table 3 (error bars are $\pm1 \sigma$).
The solid line is a weighted least-squares, linear fit.
\label{fig9}}
\end{figure}

\begin{figure}
\caption{
The variation in systemic velocity, $\gamma_1$, over
the last century.
The data are from \citet{har87} and from our solutions
in Table~3. 
\label{fig10}}
\end{figure}

\begin{figure}
\caption{
The reconstructed profiles of \ion{He}{1} $\lambda 6678$ in the
center of mass reference frame.
The profiles for the secondary ({\it middle}) and 
tertiary ({\it bottom}) are offset by $-0.2$ and $-0.4$ 
for clarity.  The tertiary is also plotted on a $3\times$ 
expanded intensity scale ({\it dotted line}). 
\label{fig11}}
\end{figure}

\begin{figure}
\caption{
The reconstructed primary star spectrum in the region $1700 - 1760$\AA~
which includes \ion{N}{4} $\lambda 1718$ and \ion{N}{3} $\lambda\lambda1748, 1751$.
The thick line shows the two component reconstruction (no tertiary flux), 
and the thin line shows the three component reconstruction in which the 
tertiary contributes 22\% of the total UV flux.
The dotted line is a spectrum of HD~209975 (19~Cep, O9.5~Ib),
a single star with a classification similar to the primary's,
which was broadened to match the rotational line broadening of the primary.
\label{fig12}}
\end{figure}

\begin{figure}
\caption{
The two component reconstructed secondary star spectrum in the same region as Fig.~12.
The dashed line shows a segmented polynomial fit of the {\it IUE} echelle ripple 
that dominates in this area of the reconstructed secondary spectrum.
The reconstruction was divided by this broad ripple function to 
yield the stellar spectral features ({\it thick solid line}). 
The dotted line shows a comparison spectrum of HD~24760 ($\epsilon$~Per, B0.5~IV-III),
a star with a classification similar to the secondary's.
\label{fig13}}
\end{figure}

\begin{figure}
\caption{
The reconstructed tertiary star spectrum in the same region as Fig.~12.
The thick line shows the featureless tertiary spectrum while 
the dotted line shows a comparison spectrum of 
the rapid rotator, HD~149757 ($\zeta$~Oph, O9.5~V, 
$V\sin i = 400 {\rm ~km~s}^{-1}$). 
\label{fig14}}
\end{figure}

\begin{figure}
\caption{
The {\it Hipparcos} light curve for $\delta$~Ori~A 
({\it open circles}) plotted against photometric phase
based upon the spectroscopically predicted time of primary eclipse
(photometric phase 0.0). 
The solid line is a model light curve for $R_p=13 R_\odot$ and $i = 67^\circ$,
and the dashed line is the same for $R_p= 9 R_\odot$ and $i = 77^\circ$.
\label{fig15}}
\end{figure}


\clearpage

\setcounter{figure}{0}

\begin{figure}
\plotone{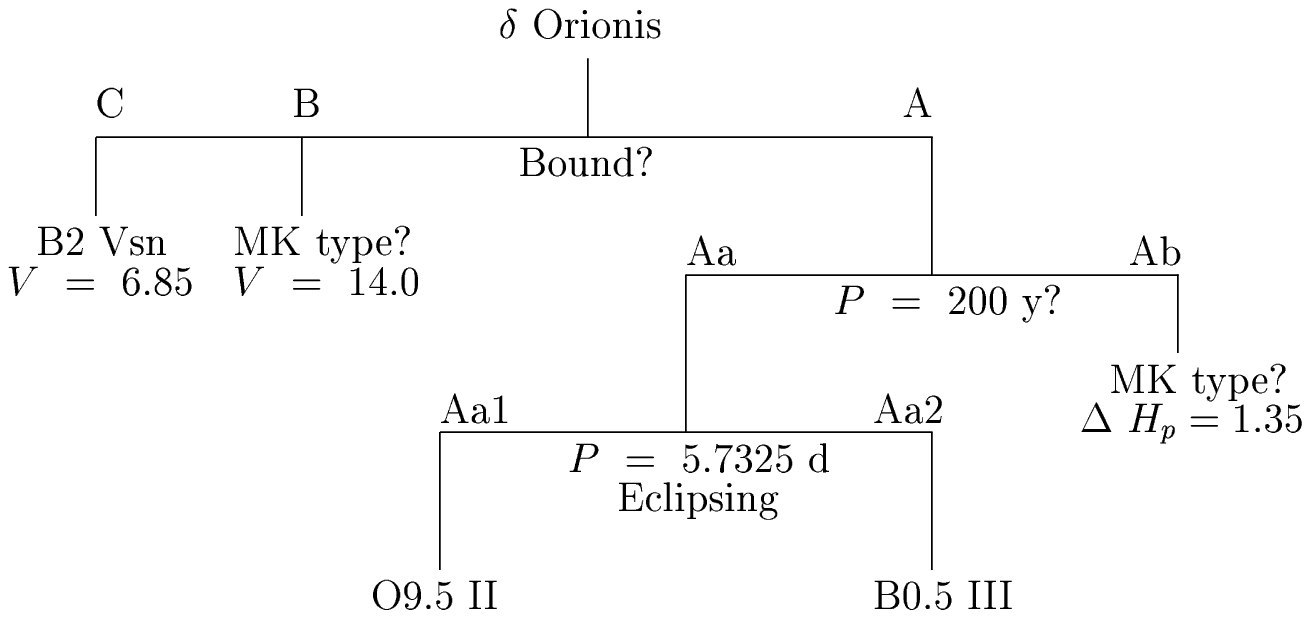}
\caption{}
\end{figure}

\begin{figure}
\plotone{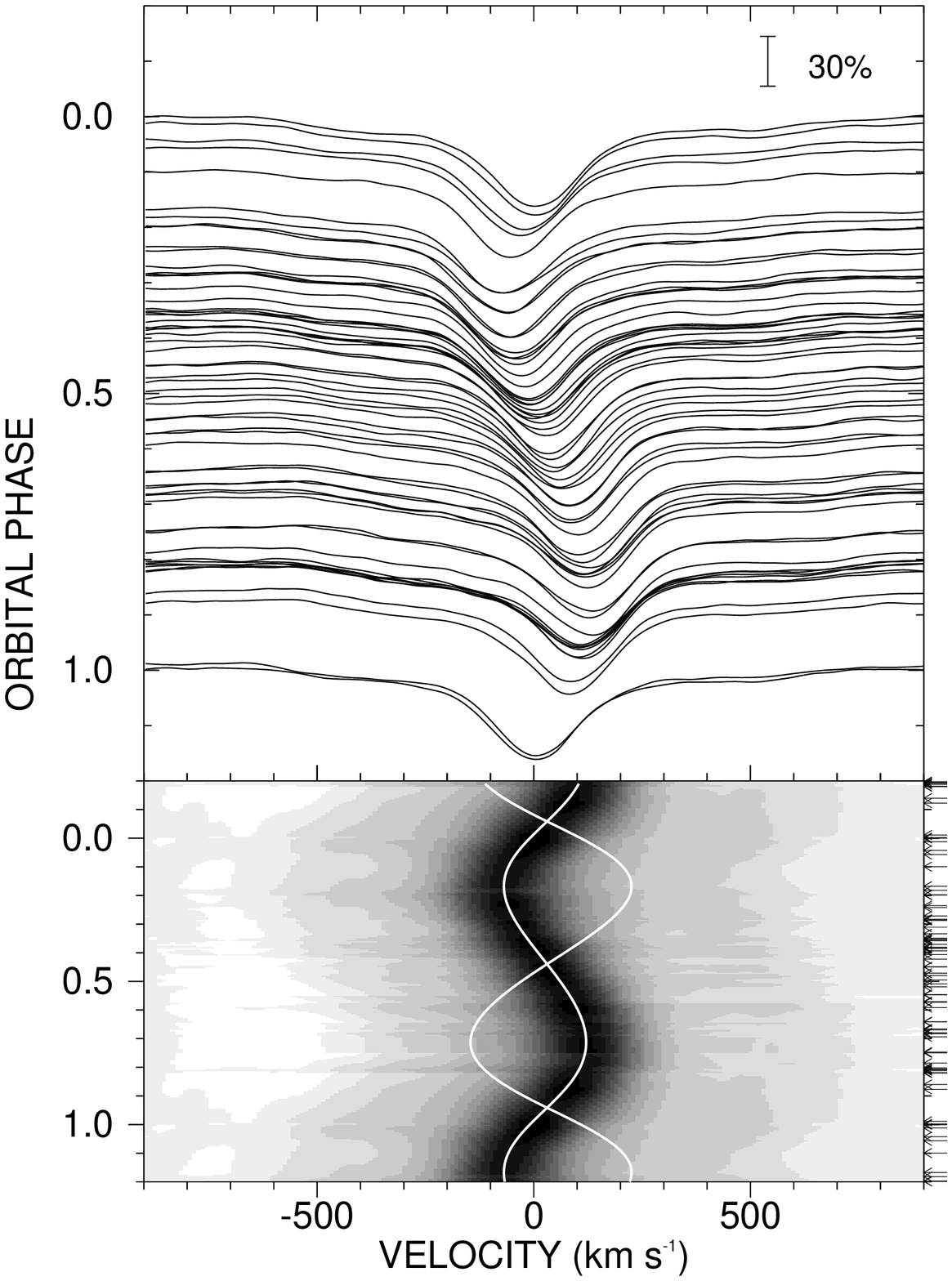}
\caption{}
\end{figure}

\begin{figure}
\plotone{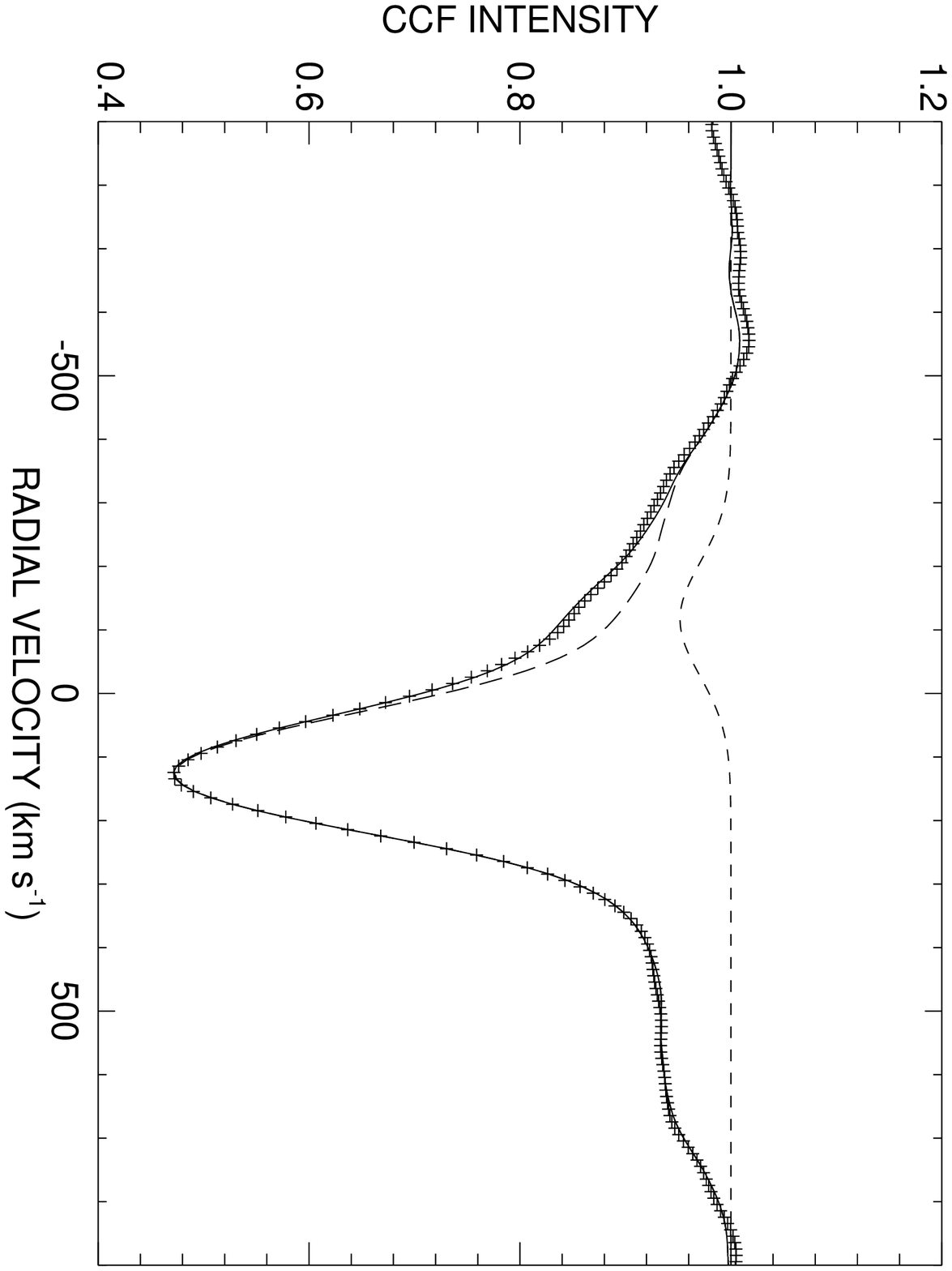}
\caption{}
\end{figure}

\begin{figure}
\plotone{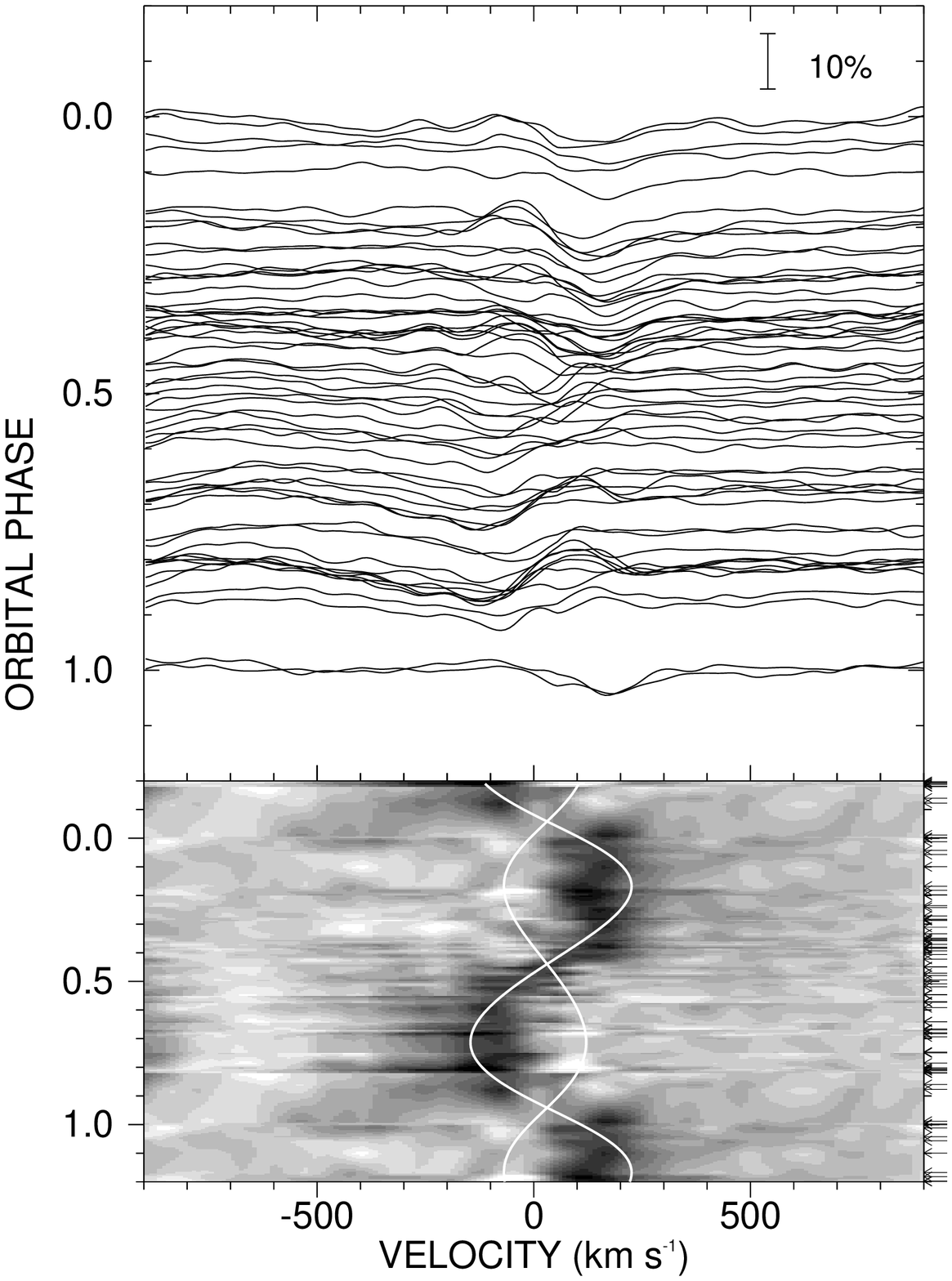}
\caption{}
\end{figure}

\begin{figure}
\plotone{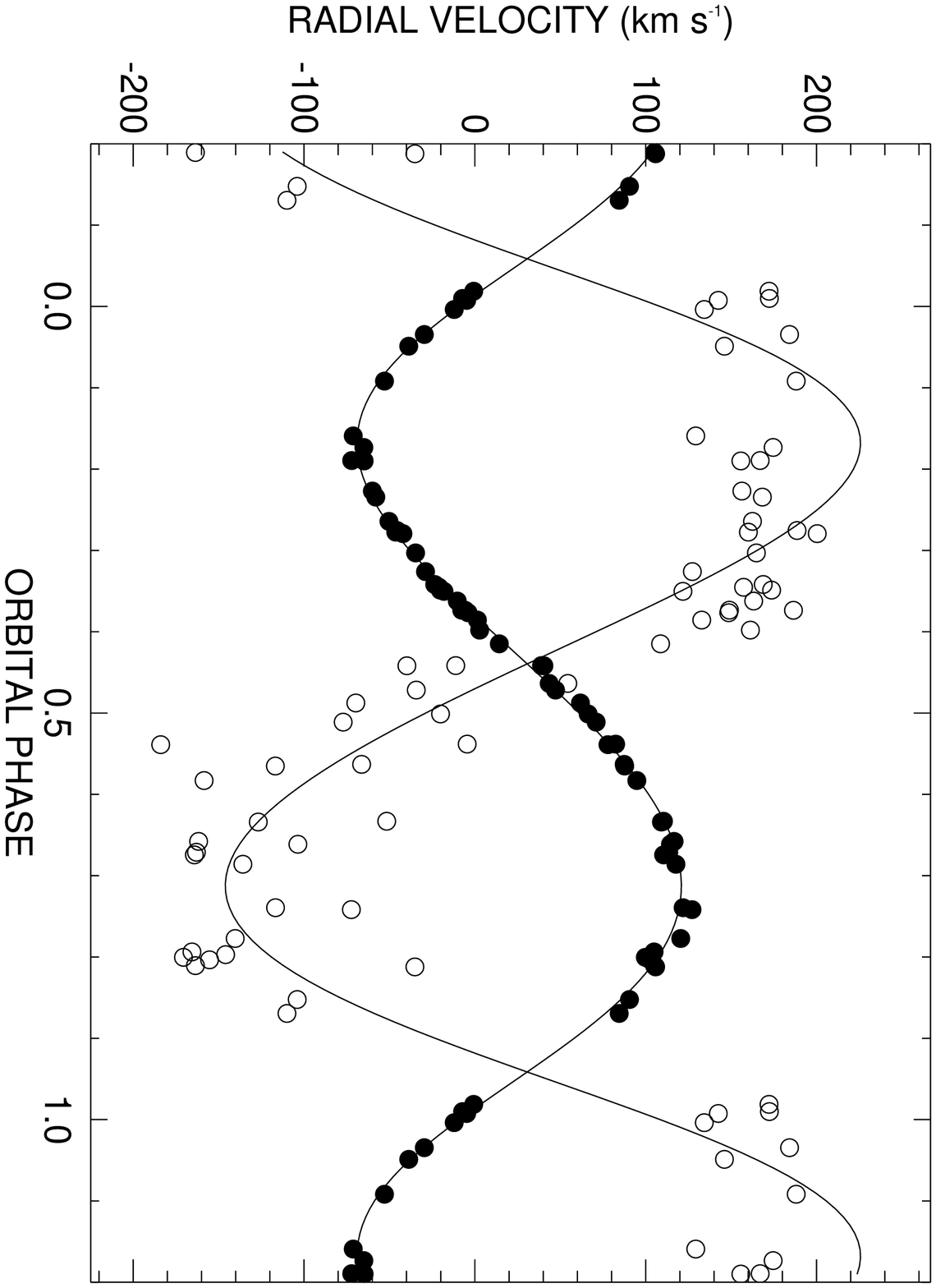}
\caption{}
\end{figure}

\begin{figure}
\plotone{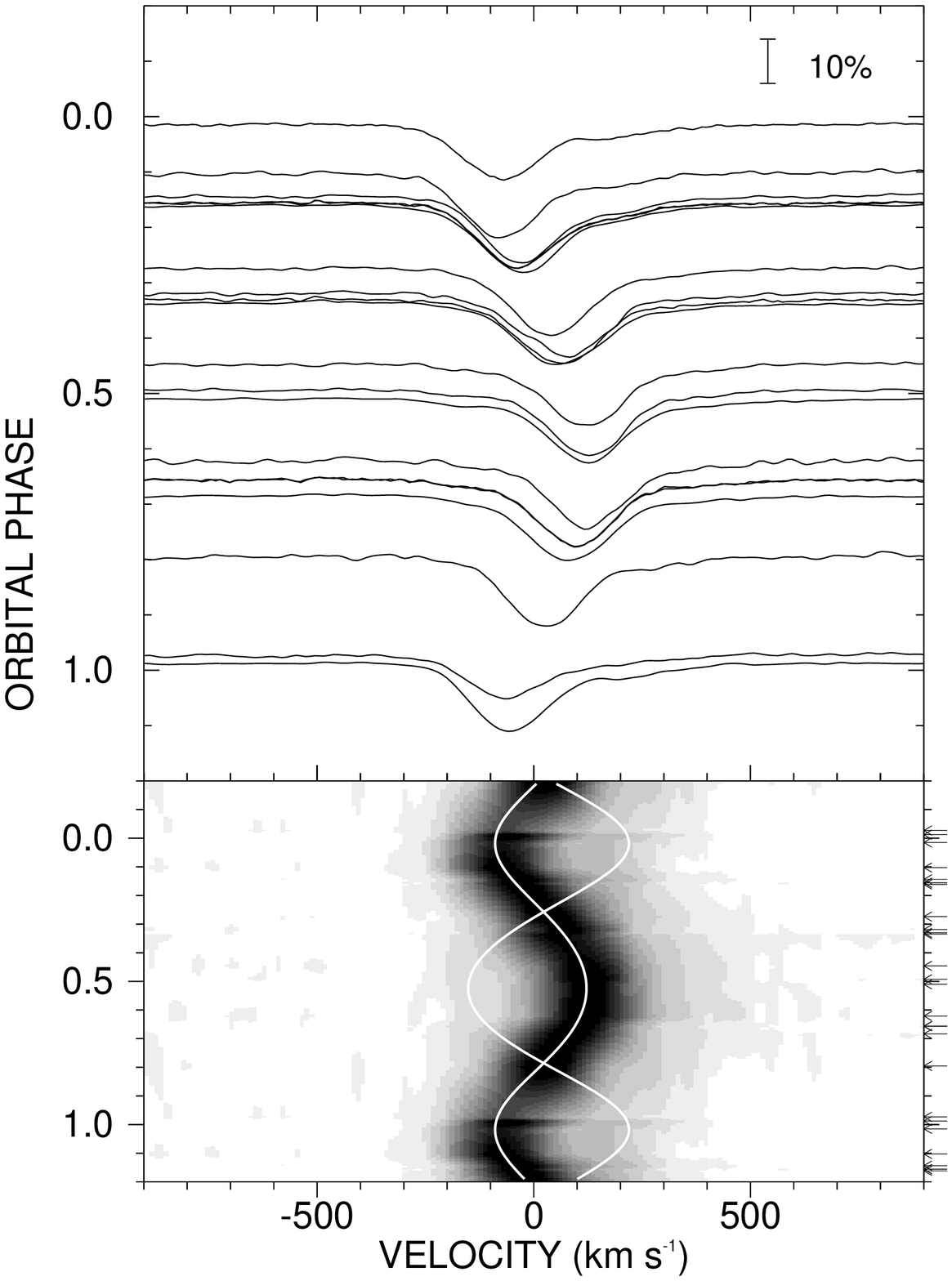}
\caption{}
\end{figure}

\begin{figure}
\plotone{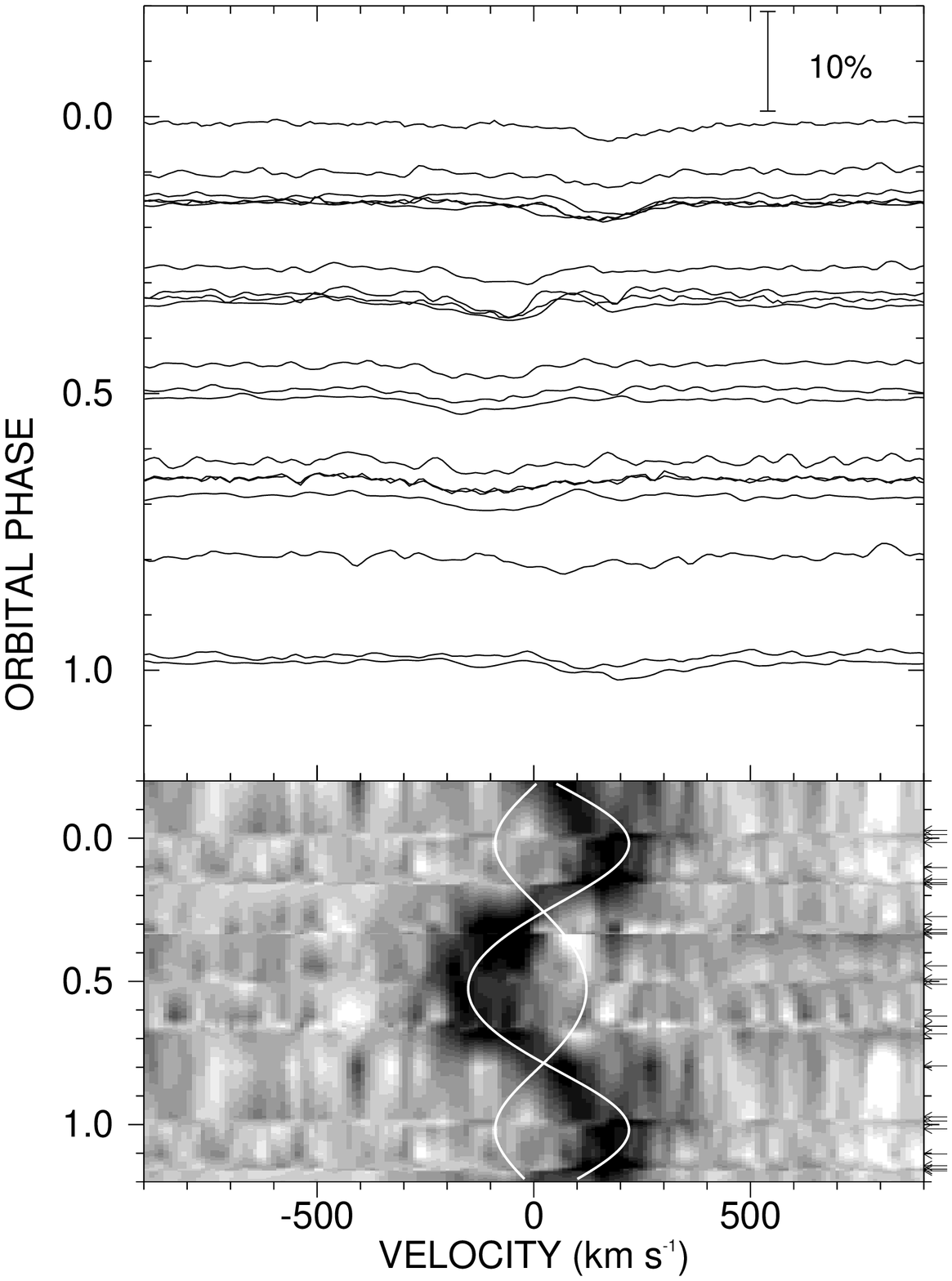}
\caption{}
\end{figure}

\begin{figure}
\plotone{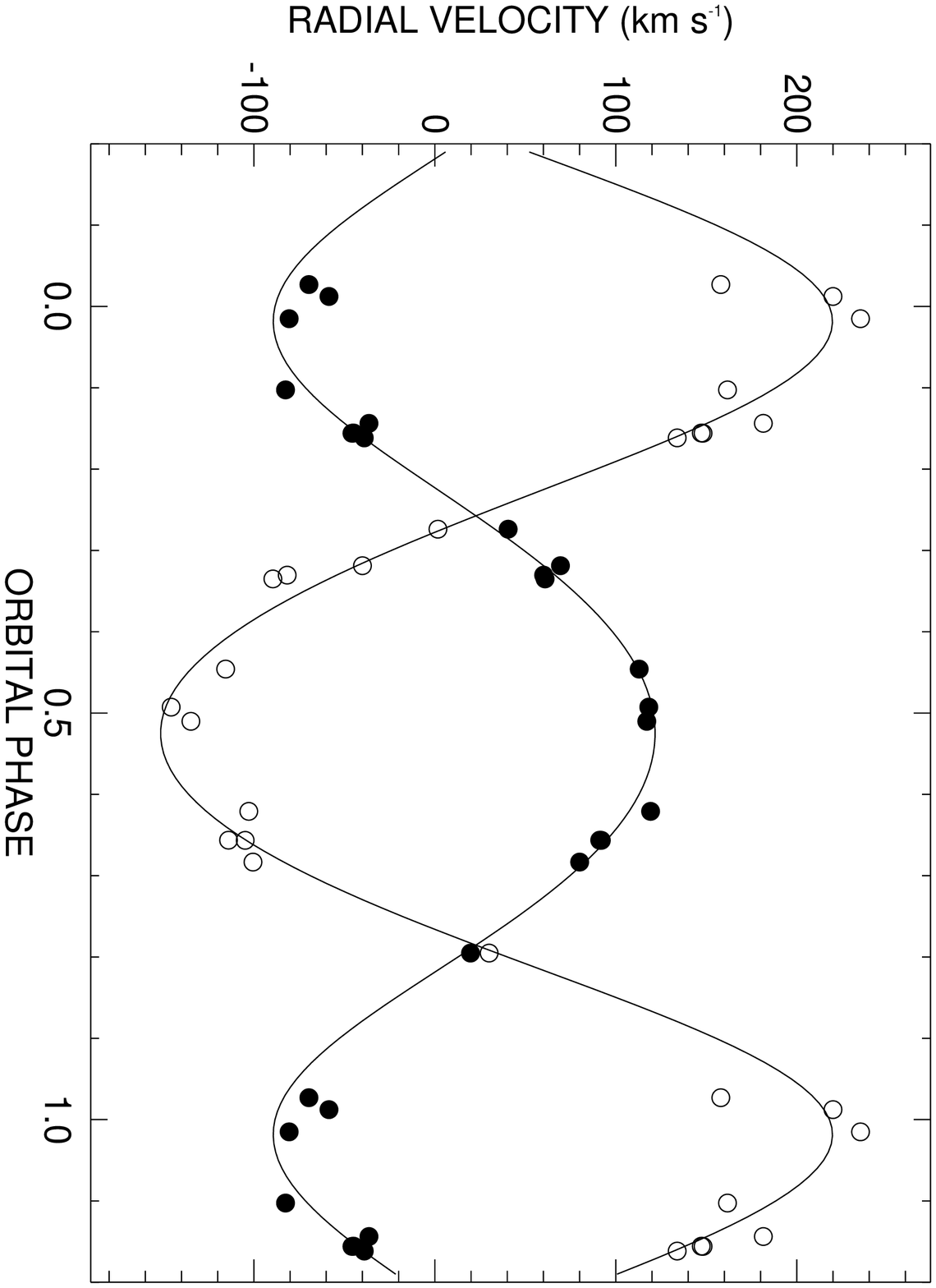}
\caption{}
\end{figure}

\begin{figure}
\plotone{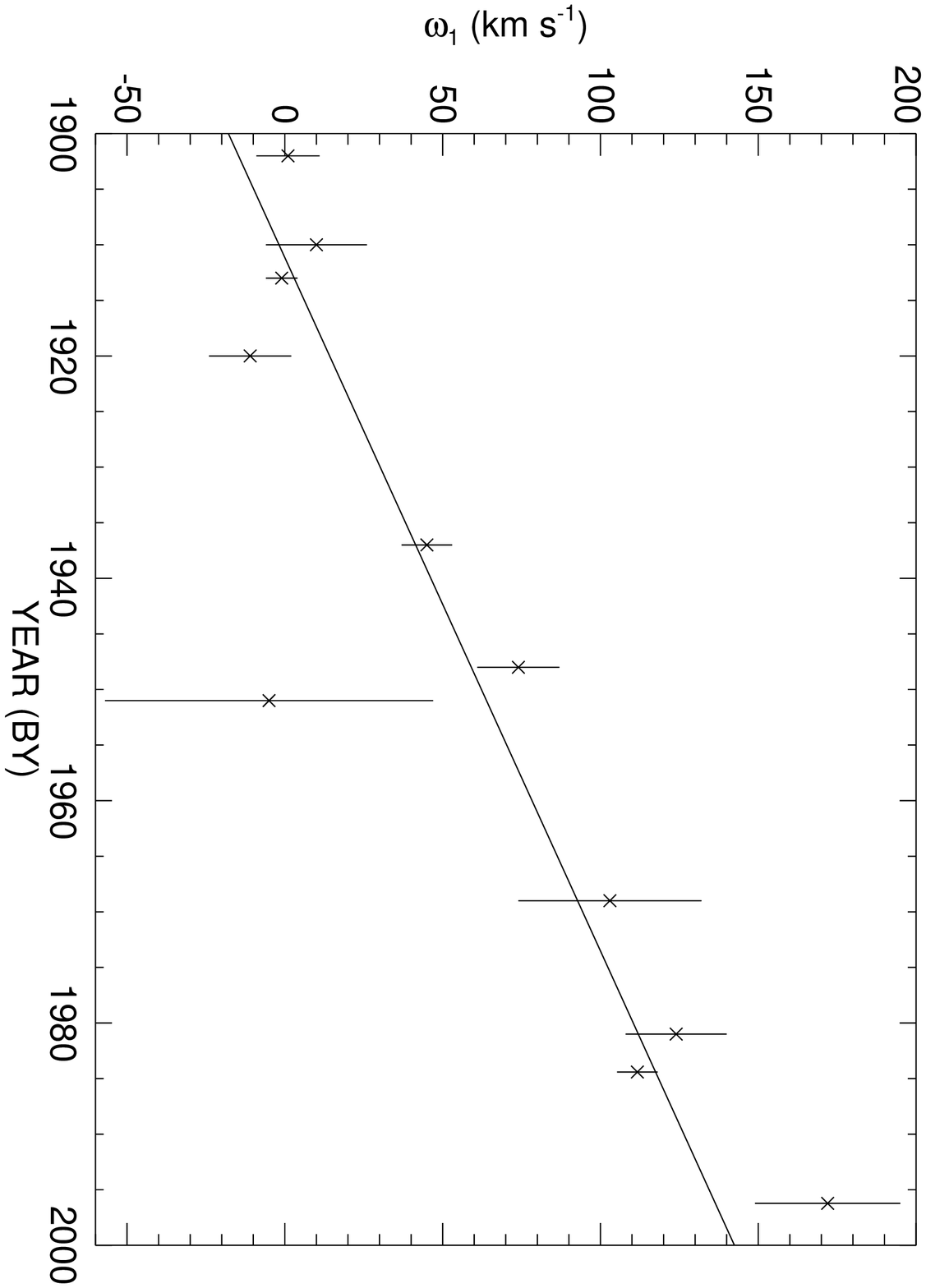}
\caption{}
\end{figure}

\begin{figure}
\plotone{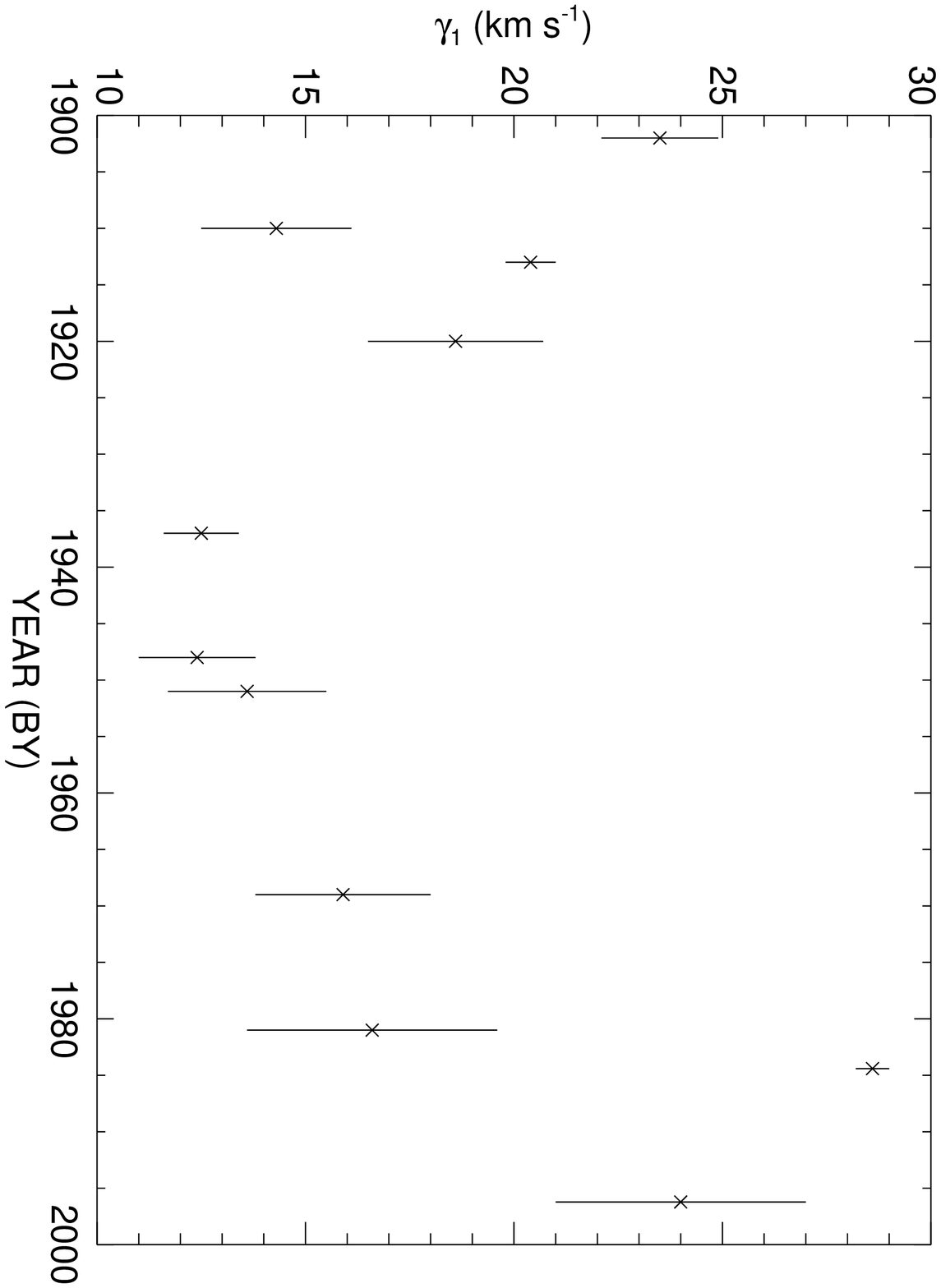}
\caption{}
\end{figure}

\begin{figure}
\plotone{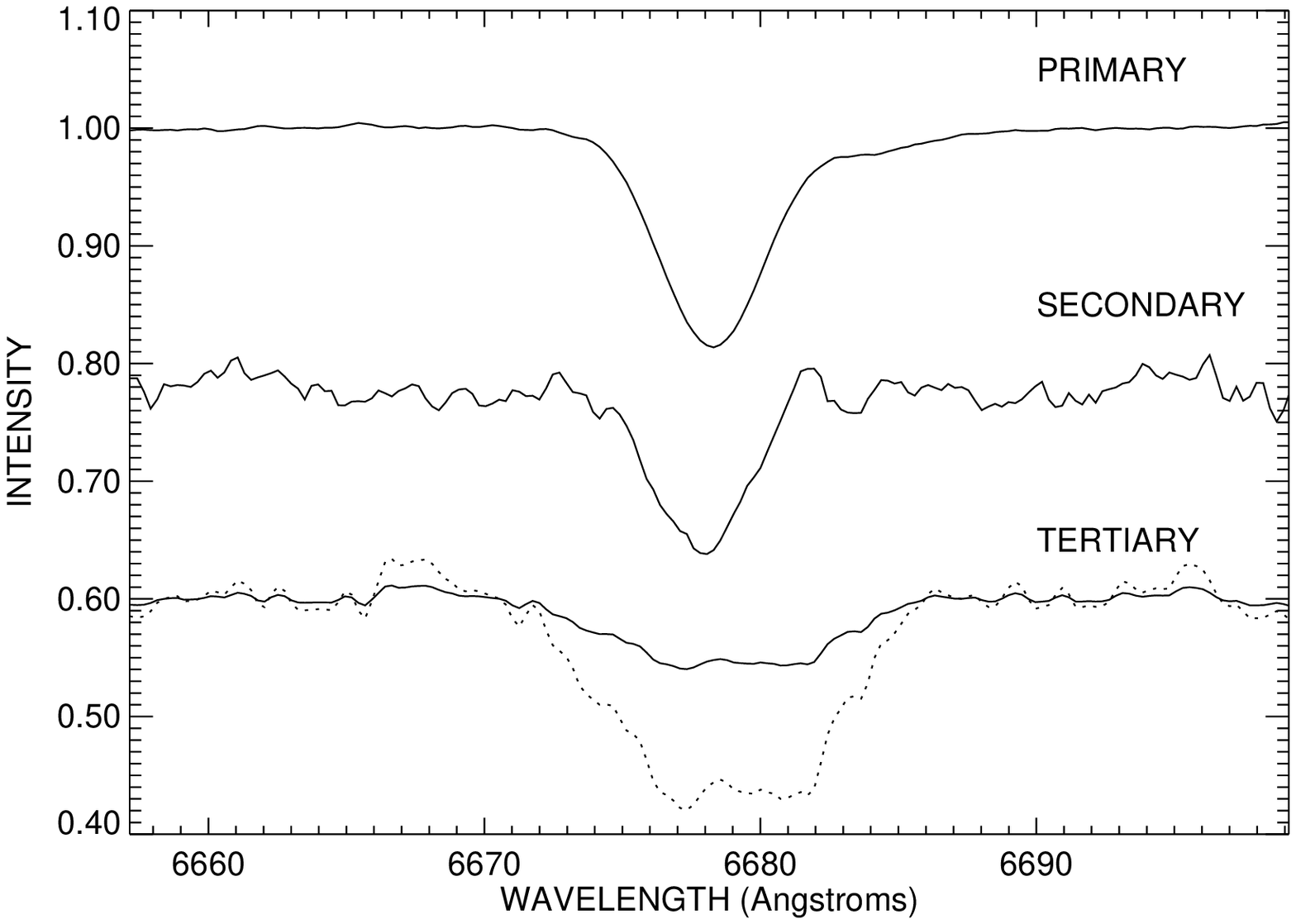}
\caption{}
\end{figure}

\begin{figure}
\plotone{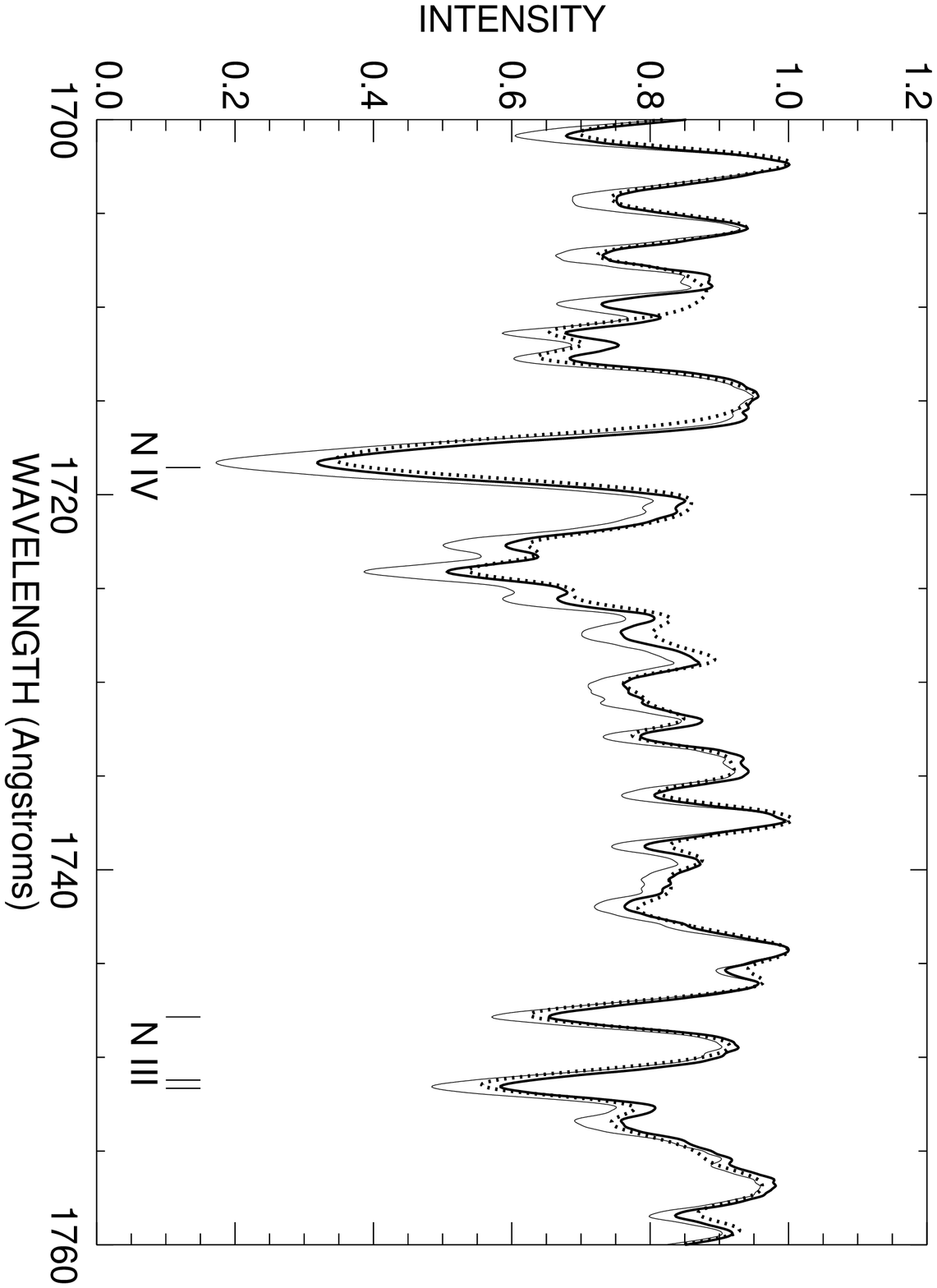}
\caption{}
\end{figure}

\begin{figure}
\plotone{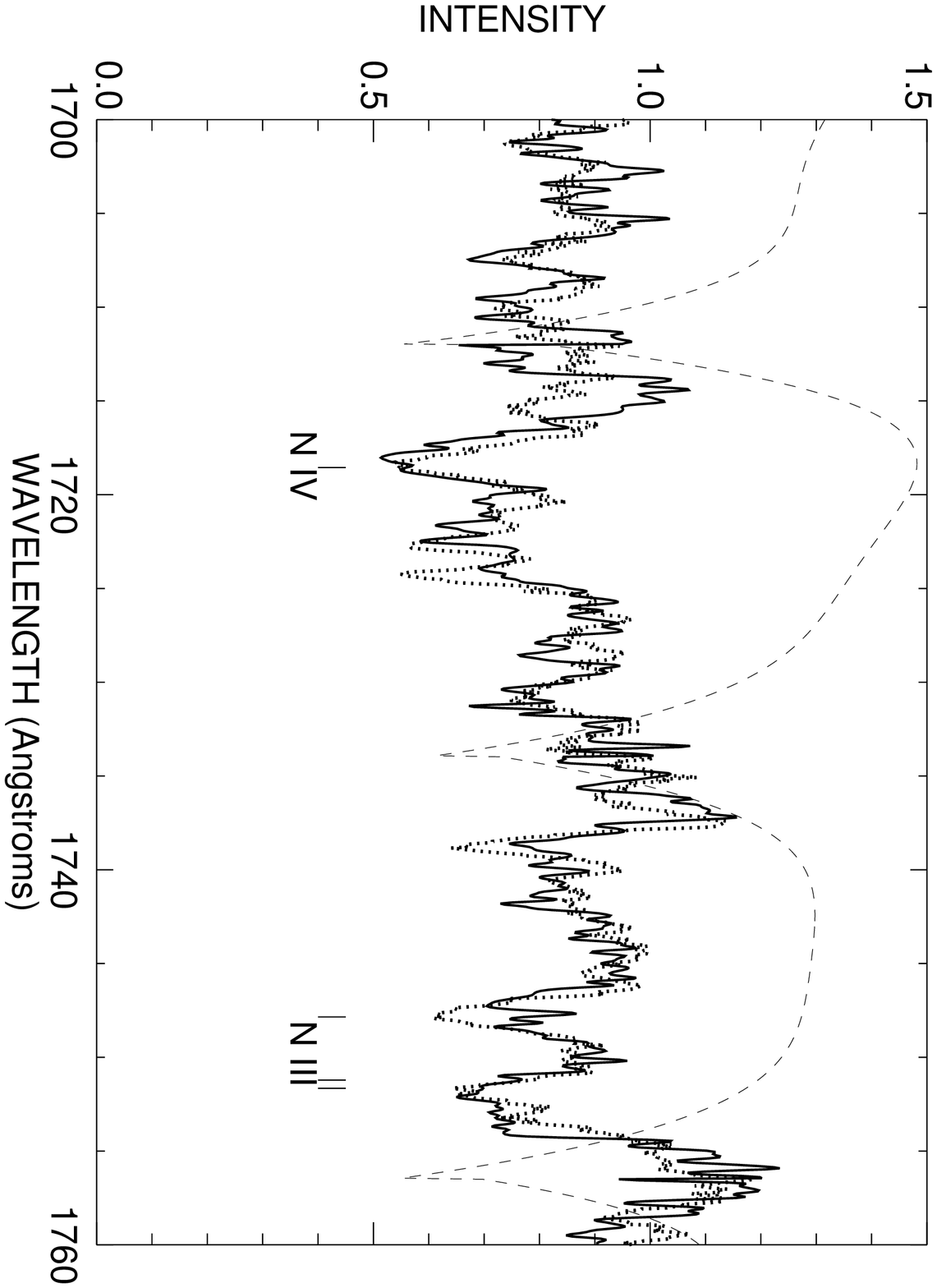}
\caption{}
\end{figure}

\begin{figure}
\plotone{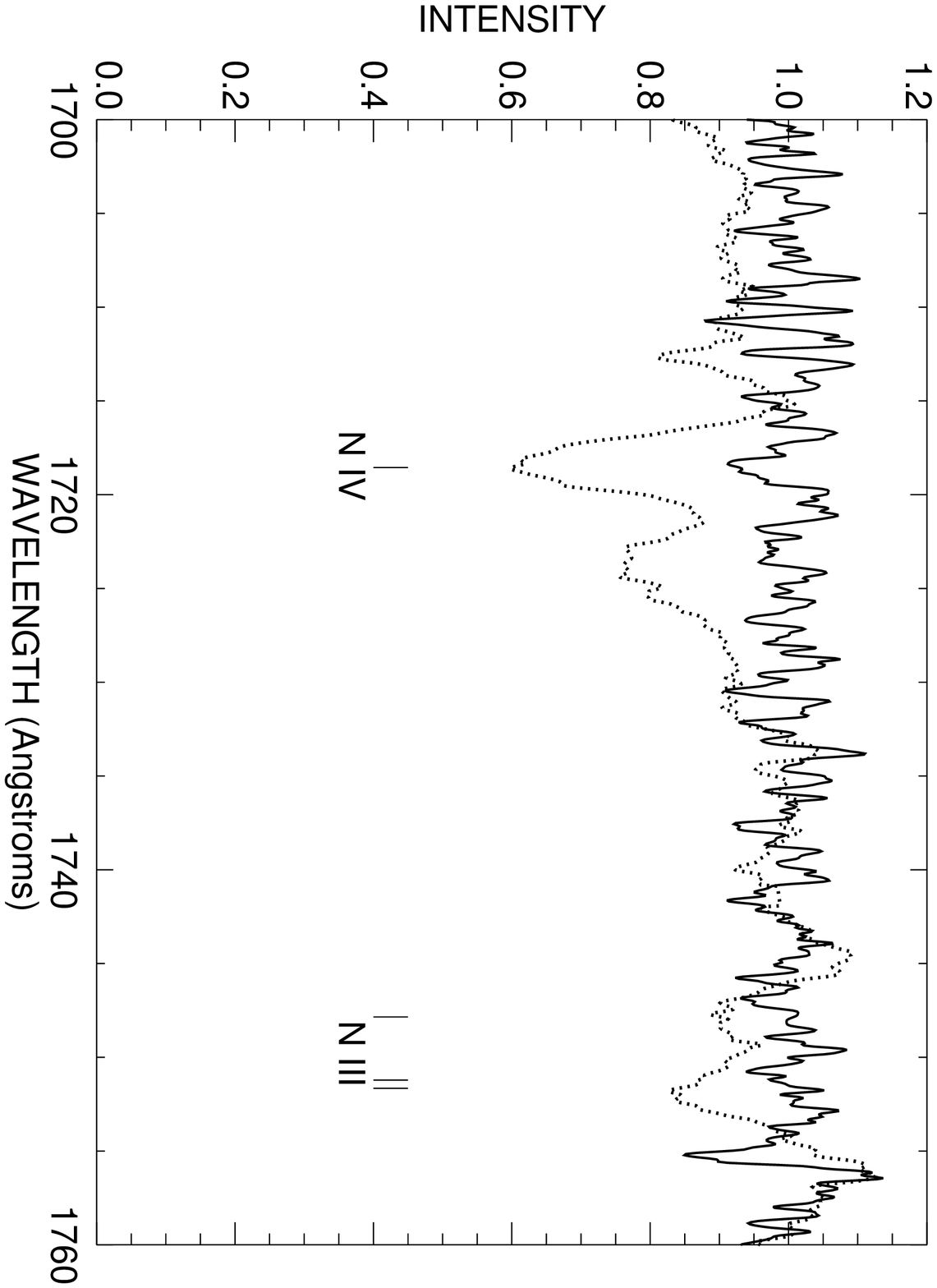}
\caption{}
\end{figure}

\begin{figure}
\plotone{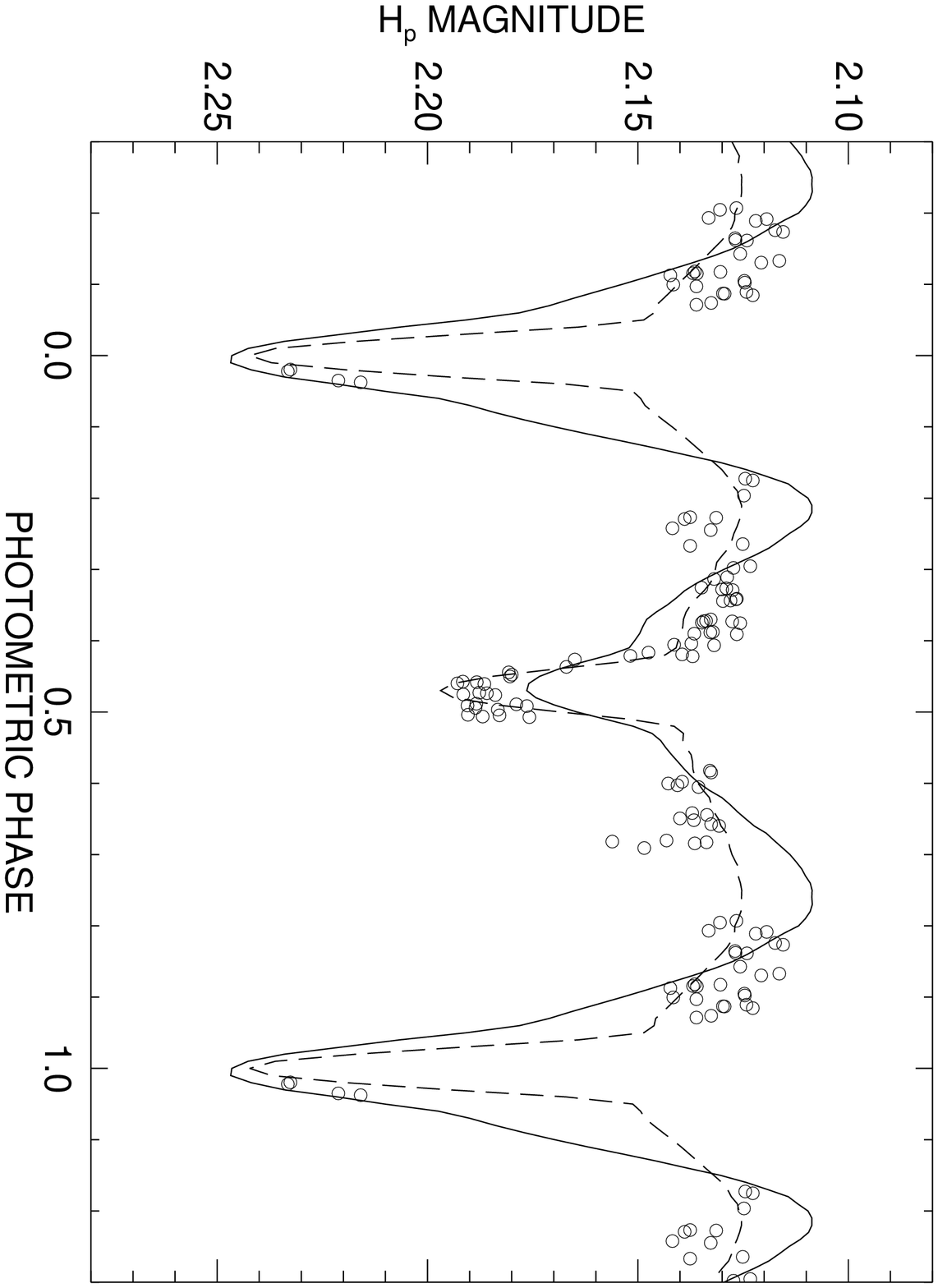}
\caption{}
\end{figure}



\begin{thebibliography}{}
\bibitem[Bagnuolo et al.(1994)]{bag94}
    Bagnuolo, W. G., Jr., Gies, D. R., Hahula, M. E.,
    Wiemker, R., \& Wiggs, M. S.
    1994, \apj, 423, 446
\bibitem[Bieging et al.(1989)]{bie89}
    Bieging, J. H., Abbott, D. C., \& Churchwell, E. B. 1989, \apj, 340, 518
\bibitem[Blaauw(1991)]{bla91}
    Blaauw, A. 1991, in The Physics of Star Formation and Early Stellar Evolution,
    NATO Advanced Science Institutes (ASI) Series C, Vol. 342, 
    ed. C. J. Lada \& N. D. Kylafis (Dordrecht: Kluwer), 125
\bibitem[Brown(1996)]{bro96}
    Brown, A. G. A. 1996, Ph.D. thesis, Univ. of Leiden 
\bibitem[Brown et al.(1994)]{bro94}
    Brown, A. G. A., de Geus, E. J., \& de Zeeuw, P. T. 1994, \aap, 289, 101
\bibitem[Curtiss(1914)]{cur14}
    Curtiss, R. H. 1914, Publ. Mich. Obs., 1, 118
\bibitem[de Zeeuw et al.(1999)]{dez99}
    de Zeeuw, P. T., Hoogerwerf, R., de Bruijne, J. H. J., Brown, A. G. A., \& Blaauw, A.
    1999, \aj, 117, 354
\bibitem[Fullerton(1990)]{ful90}
    Fullerton, A. W. 1990, Ph.D. thesis, Univ. of Toronto
\bibitem[Galkina(1976)]{gal76}
    Galkina, T. S. 1976, Izvestiia Krymskaia Astrofiz. Obs., 54, 128
\bibitem[Garhart et al.(1997)]{gar97}
    Garhart, M. P., Smith, M. A., Turnrose, B. E., Levay, K. L., \& Thompson, R. W.
    1997, International Ultraviolet Explorer New Spectral Image Processing System
    Information Manual, Version 2.0 (Greenbelt: NASA)
\bibitem[Gies(1995)]{gie95}
    Gies, D.~R.~1995, 
    in IAU Symposium 163, Wolf-Rayet Stars: Binaries, Colliding Winds,
    Evolution, ed.\ K.\ A.\ van der Hucht \& P.\ M.\ Williams
    (Dordrecht: Kluwer), 373
\bibitem[Gies \& Bolton(1986)]{gie86}
    Gies, D. R., \& Bolton, C. T. 1986, \apjs, 61, 419
\bibitem[Gies et al.(1999)]{gie99}
    Gies, D. R., et al. 1999, \apj, 525, 420
\bibitem[Harmanec(1988)]{har88}
    Harmanec, P. 1988, Bull. Ast. Inst. Czech., 39, 329
\bibitem[Harmanec(1998)]{hrm98}
    Harmanec, P. 1998, \aap, 335, 173
\bibitem[Harries et al.(1998)]{har98}
    Harries, T. J., Hilditch, R. W., \& Hill, G. 1998, \mnras, 295, 386
\bibitem[Hartkopf et al.(1993)]{har93}
    Hartkopf, W. I., Mason, B. D., Barry, D. J., McAlister, H. A., Bagnuolo, W. G., Jr.,
    \& Prieto, C. M. 1993, \aj, 106, 352
\bibitem[Hartmann(1904)]{har04}
    Hartmann, J. 1904, \apj, 19, 268
\bibitem[Harvey et al.(1987)]{har87}
    Harvey, A. S., Stickland, D. J., Howarth, I. D., \& Zuiderwijk, E. J. 
    1987, Observatory, 107, 205
\bibitem[Heger \& Langer(2000)]{heg00}
    Heger, A., \& Langer, N. 2000, \apj, 544, 1016
\bibitem[Heintz(1980)]{hei80}
    Heintz, W. D. 1980, \apjs, 44, 111
\bibitem[Herrero et al.(1999)]{her99}
    Herrero, A., Corral, L. J., Villamariz, M. R., \& Martin, E. L. 1999, \aap, 348, 542
\bibitem[Herrero et al.(2000)]{her00}
    Herrero, A., Puls, J., \& Villamariz, M. R., 2000, \aap, 354, 193
\bibitem[Hnatek(1920)]{hna20}
    Hnatek, A. 1920, Astronomisches Nachrichten, 213, 17
\bibitem[Howarth \& Prinja(1989)]{how89}
    Howarth, I. D., \& Prinja, R. K. 1989, \apjs, 69, 527
\bibitem[Howarth et al.(1997)]{how97}
    Howarth, I. D., Siebert, K. W., Hussain, G. A. J., \& Prinja, R. K. 1997, \mnras, 284, 265
\bibitem[Hutchings(1976)]{hut76}
    Hutchings, J. B. 1976, \apj, 203, 438
\bibitem[Jordan(1914)]{jor14}
    Jordan, F. C. 1914, Publ. Allegheny Obs., 3, 125
\bibitem[Koch \& Hrivnak(1981)]{koc81}
    Koch, R. H., \& Hrivnak, B. J. 1981, \apj, 248, 249
\bibitem[Kurucz(1994)]{kur94}
    Kurucz, R.~L.\ 1994, Solar Abundance Model Atmospheres for
    0, 1, 2, 4, 8 km/s, Kurucz CD-ROM No.\ 19 (Cambridge, MA:
    Smithsonian Astrophysical Obs.)
\bibitem[Liu et al.(1997)]{liu97}
    Liu, N., et al. 1997, \apj, 485, 350
\bibitem[Luyten et al.(1939)]{luy39}
    Luyten, W. J., Struve, O., \& Morgan, W. W. 1939, Publ. Yerkes Obs., 7, 256
\bibitem[Mason et al.(1998)]{mas98}
    Mason, B. D., Gies, D. R., Hartkopf, W. I., Bagnuolo, W. G., Jr., ten Brummelaar, T., \&
    McAlister, H. A. 1998, \aj, 115, 821
\bibitem[Massa(1989)]{mas89}
    Massa, D. 1989, \aap, 224, 131
\bibitem[McAlister et al.(1983)]{mca83}
    McAlister, H. A., Hartkopf, W. I., Hendry, E. M., Campbell, B. G., \& Fekel, F. C.
    1983, \apjs, 51, 309
\bibitem[McAlister et al.(1987)]{mca87}
    McAlister, H. A., Hartkopf, W. I., Hutter, D. J., \& Franz, O. G. 1987, \aj, 93, 688
\bibitem[McAlister et al.(1989)]{mca89}
    McAlister, H. A., Hartkopf, W. I., Sowell, J. R., Dombrowski, E. G., \& Franz, O. G.
    1989, \aj, 97, 510
\bibitem[McAlister \& Hendry(1982)]{mca82}
    McAlister, H. A., \& Hendry, E. M. 1982, \apjs, 49, 267
\bibitem[McAlister et al.(1993)]{mca93}
    McAlister, H. A., Mason, B. D., Hartkopf, W. I., \& Shara, M. M. 1993, \aj, 106, 1639
\bibitem[Meynet \& Maeder(2000)]{mey00}
    Meynet, G., \& Maeder, A. 2000, \aap, 361, 101
\bibitem[Miczaika(1952)]{mic52}
    Miczaika, G. R. 1952, Zeitschrift fur Astrophysik, 30, 299
\bibitem[Mochnacki \& Doughty(1972)]{moc72}
    Mochnacki, S.~W., \& Doughty, N.~A.~1972, \mnras, 156, 51
\bibitem[Monet(1980)]{mon80}
    Monet, D. G. 1980, \apj, 237, 513
\bibitem[Morbey \& Brosterhus(1974)]{mor74}
    Morbey, C. L., \& Brosterhus, E. B. 1974, \pasp, 86, 455
\bibitem[Natarajan \& Rajamohan(1971)]{nat71}
    Natarajan, V., \& Rajamohan, R. 1971, Kodaikanal Bull., 208
\bibitem[Penny(1996)]{pen963}
    Penny, L. R. 1996, \apj, 463, 737
\bibitem[Penny et al.(1997)]{pen97}
    Penny, L. R., Gies, D. R., \& Bagnuolo, W. G., Jr.
    1997, \apj, 483, 439
\bibitem[Penny et al.(1999)]{pen99}
    Penny, L. R., Gies, D. R., \& Bagnuolo, W. G., Jr.
    1999, \apj, 518, 450
\bibitem[Penny et al.(2001)]{pen01}
    Penny, L. R., et al. 2001, \apj, 548, 889
\bibitem[Perryman(1997)]{per97}
    Perryman, M. A. C. 1997,
    The \emph{Hipparcos} and \emph{Tycho} Catalogues (ESA SP-1200) (Noordwijk: ESA)
\bibitem[Pismis et al.(1950)]{pis50}
    Pismis, P., Haro, G., \& Struve, O. 1950, \apj, 111, 509
\bibitem[Prieur et al.(2001)]{pri01}
    Prieur, J.-L., Oblak, E., Lampens, P., Kurpinska-Winiarska, M., Aristidi, E.,
    Koechlin, L., \& Ruymaekers, G. 2001, \aap, 367, 865
\bibitem[Reid et al.(1993)]{rei93}
    Reid, A. H. N., et al. 1993, \apj, 417, 320
\bibitem[Rountree \& Sonneborn(1993)]{rou93}
    Rountree, J., \& Sonneborn, G. 1993,
    Spectral Classification with the International Ultraviolet Explorer:
    An Atlas of B-type Spectra (NASA RP-1312) (Washington, DC: NASA)
\bibitem[Runacres \& Bloome(1996)]{run96}
    Runacres, M. C., \& Bloome, R. 1996, \aap, 309, 544
\bibitem[Schaller et al.(1992)]{sch92}
    Schaller, G., Schaerer, D., Meynet, G., \&  Maeder, A. 1992, \aaps, 96, 269
\bibitem[Sch\"{o}nberner \& Harmanec(1995)]{sch95}
    Sch\"{o}nberner, D., \& Harmanec, P. 1995, \aap, 294, 509
\bibitem[Singh(1982)]{sin82}
    Singh, M. 1982, \apss, 87, 269
\bibitem[Tarasov et al.(1995)]{tar95}
    Tarasov, A. E., et al. 1995, A\&AS, 110, 59
\bibitem[Thaller(1997a)]{tha97a}
    Thaller, M. L. 1997a, \apj, 487, 380
\bibitem[Thaller(1997b)]{tha97b}
    Thaller, M. L. 1997b, Ph.D. dissertation, Georgia State Univ. 
\bibitem[Vanbeveren et al.(1998)]{van98}
    Vanbeveren, D., van Rensbergen, W., \& de Loore, C. 1998,
    The Brightest Binaries (Dordrecht: Kluwer)
\bibitem[Voels et al.(1989)]{voe89}
    Voels, S. A., Bohannan, B., Abbott, D. C., \& Hummer, D. G. 1989, \apj, 340, 1073
\bibitem[Wade \& Rucinski(1985)]{wad85}
    Wade, R.~A., \& Rucinski, S.~M. 1985, A\&AS, 60, 471
\bibitem[Walborn(1972)]{wal72}
    Walborn, N. R. 1972, \aj, 77, 312
\bibitem[Walborn(1976)]{wal76}
    Walborn, N. R. 1976, \apj, 205, 419
\bibitem[Wellstein et al.(2001)]{wel01}
    Wellstein, S., Langer, N., \& Braun, H. 2001, \aap, 369, 939
\bibitem[Worley \& Douglass(1997)]{wor97}
    Worley, C. E., \& Douglass, G. G. 1997, \aaps, 125, 523
\end{thebibliography}
\end{document}